\documentclass[11pt,letterpaper]{article}
\usepackage[T1]{fontenc}
\usepackage{lmodern}
\usepackage[margin=1in]{geometry}
\usepackage{amsmath,amssymb,amsthm,mathtools}
\usepackage{microtype,booktabs}
\usepackage[numbers,sort&compress]{natbib}
\usepackage{xurl}
\usepackage{xcolor}
\definecolor{PaperBlue}{RGB}{25,62,125}
\definecolor{PaperGreen}{RGB}{15,105,80}
\definecolor{PaperPurple}{RGB}{105,45,120}
\usepackage[colorlinks=true,linkcolor=PaperBlue,citecolor=PaperGreen,urlcolor=PaperPurple,linktoc=all,breaklinks=true]{hyperref}
\usepackage{bookmark}
\hypersetup{pdftitle={Exact Quantum Algorithms for Composed Boolean Functions},
  pdfsubject={Exact quantum algorithms, composition, recursive functions, and separations}}
\numberwithin{equation}{section}
\newtheorem{theorem}{Theorem}[section]
\newtheorem{lemma}[theorem]{Lemma}
\newtheorem{proposition}[theorem]{Proposition}
\newtheorem{corollary}[theorem]{Corollary}
\newtheorem{conjecture}[theorem]{Conjecture}
\theoremstyle{definition}

\newtheorem{problem}[theorem]{Problem}
\newtheorem{remark}[theorem]{Remark}
\newcommand{\proofstep}[1]{\par\smallskip\noindent\textbf{#1}\enspace\ignorespaces}
\newcommand{\QE}{Q_E}
\newcommand{\Adv}{\operatorname{Adv}^{\pm}}
\newcommand{\OR}{\mathrm{OR}}
\newcommand{\AND}{\mathrm{AND}}
\newcommand{\NAE}{\mathrm{NAE}}
\newcommand{\PAR}{\mathrm{PARITY}}
\newcommand{\eps}{\varepsilon}
\newcommand{\QElim}{Q_E^{\lim}}
\newcommand{\bs}{\operatorname{bs}}
\newcommand{\fbs}{\operatorname{fbs}}
\newcommand{\cert}{\operatorname{C}}
\newcommand{\ket}[1]{\lvert #1\rangle}
\newcommand{\bra}[1]{\langle #1\rvert}
\title{Exact Quantum Algorithms for Composed Boolean Functions}

\author{ Gagan Chiruvella\footnote{Indian Institute of Technology Hyderabad. Email: {\tt cs24btech11020@iith.ac.in}}
\and Chandrima Kayal\footnote{Universit\'e Paris Cit\'e, CNRS, IRIF, Paris, France, Email: {\tt chandrimakayal2012@gmail.com}}
 \and Rajat Mittal\footnote{Indian Institute of Technology Kanpur. Email: {\tt rmittal@cse.iitk.ac.in}}
 \and
 Manaswi Paraashar\footnote{Indian Institute of Technology Hyderabad. Email: {\tt manaswi@cse.iith.ac.in}}
 \and
 Nitin Saurabh\footnote{Indian Institute of Technology Hyderabad. Email: {\tt nitin@cse.iith.ac.in}}
}
\date{}

\begin{document}
\maketitle
\vspace{-1em}
\begin{abstract}
We study exact quantum algorithms for block compositions of total Boolean
functions. For an outer function $f$ of degree $D$ and an inner function
$g$ on $m$ bits, we prove
\[
 Q_E(f\circ g)=O\!\left(mD+Q_E(f)Q_{1/(4D)}(g)\right).
\]
The algorithm combines approximate inner evaluations with a finite exact
correction. Its two costs are governed by different parameters of the
outer function: exact query complexity determines the cost of simulation,
whereas polynomial degree determines the number of inner blocks used in
the correction. 

We first establish the construction with an
arity-dependent error threshold and then obtain the degree-dependent
threshold above. 
For recursive ternary not-all-equal, this gives an exact algorithm using
$(3/\sqrt2)^{k(1+o(1))}$ queries asymptotically. 
More generally, for every 
fixed nonconstant total gadget $h$, we determine the exact recursive rate
as $\max\{\deg(h),\operatorname{Adv}^{\pm}(h)\}$ and prove
$Q_E(h^{\circ k})=\widetilde O_h(R(h^{\circ k})^2)$.

The composition limit also allows to transfer separations from $Q_{\varepsilon}$ and $\deg$ to exact quantum query complexity $Q_E$ (similar to fractional block sensitivity to block sensitivity transfer in Tal and Kulkarni 2016, Chicago Journal of Theoretical Computer Science). 
We construct a family with
$s=\operatorname{bs}=\operatorname{fbs}\ge Q_E^{3/2-o(1)}$,
improving the previous exponent
$1.1527\ldots$. A second construction gives
$R \ge Q_E^{2-o(1)}$,
improving the previous exponent $3/2$. We also allow different inner
functions and give analogous quantitative separations in quantum
communication.
\end{abstract}

\vfill

\paragraph{AI statement.} We used GPT-6 Astra and GPT-6.1 Sol to explore proof strategies, search for related literature, and revise the manuscript. 
The main proof ideas were generated by these systems when the authors used them to understand the recent work \cite{PP}.
The authors explored these ideas and related questions further.
The authors take full responsibility for the correctness and content of the paper. 

\clearpage
\setcounter{tocdepth}{2}
\begingroup
\small
\makeatletter
\renewcommand{\l@section}[2]{%
  \ifnum\c@tocdepth>\z@
    \addpenalty{\@secpenalty}\vskip .35em
    \@dottedtocline{1}{0em}{1.5em}{\bfseries #1}{\bfseries #2}%
  \fi}
\makeatother
\tableofcontents
\endgroup
\clearpage

\section{Introduction}
\label{sec:introduction}
Composition is a basic way to construct query algorithms and create examples of separations between complexity measures of Boolean functions. Given $f:\{0,1\}^n\to\{0,1\}$ and
$g:\{0,1\}^m\to\{0,1\}$, the block composition $f\circ g$ applies
$g$ to $n$ disjoint input blocks and then applies $f$ to their outputs.
An exact quantum algorithm can evaluate this composition by replacing
each query to an input bit of $f$ with an exact computation of $g$.
This simulation uses both algorithms as black boxes, but it also pays
for an exact inner computation at every outer query. Is this algorithm the best possible (in terms of number of queries used), or can we improve it? In other words, is $$Q_E(f\circ g)=\Theta \left( Q_E(f)Q_E(g) \right).$$

For comparison, degree composes exactly, and bounded-error quantum query
complexity satisfies $Q(f\circ g)=\Theta(Q(f)Q(g))$ by the general
adversary characterization and its composition theorem
\cite{Tal,Reichardt,LMRSS}. The standard bounds for exact quantum queries are
\begin{align}
 Q_E(f\circ g)&\le2Q_E(f)Q_E(g),
 \label{eq:intro-standard-upper}\\
 Q_E(f\circ g)&=\Omega\!\left(\max\{Q_E(f),Q_E(g),
       \deg(f)\deg(g),Q(f)Q(g)\}\right).
 \label{eq:intro-standard-lower}
\end{align}
Here and below the functions are nonconstant and total. The upper bound
uses coherent compute--phase--uncompute simulation; the lower bounds
follow from restrictions, the polynomial method, and adversary
composition~\cite{BBCMW,Reichardt,LMRSS}. Pednekar and
Podder~\cite{PP} showed that exact quantum complexity need not compose
multiplicatively. In particular, they gave the bound
\[
 Q_E(f\circ\OR_m)=O\!\left(m\deg(f)+n^2\sqrt m\right)
\]
and used it to construct total functions with
$Q_E(f\circ g)=o(Q_E(f)Q_E(g))$.

We ask whether approximate inner computations can reduce $Q_E(f \circ g)$ in general, while the final answer remains exact. Ordinary error reduction does not resolve
this question: any nonzero probability of an incorrect answer is
unacceptable. Instead, our algorithm corrects the approximate computation
by an exact algebraic identity. The number of blocks involved in a
correction is bounded by the polynomial degree of the outer function,
which can be much smaller than its exact query complexity. Our construction applies to arbitrary total inner functions and separates
the cost of simulating the outer algorithm from the cost of exact
correction.


\subsection{Our results}
Write $Q_\eps$ for quantum query complexity with worst-case error at
most $\eps$, $Q=Q_{1/3}$, and $R=R_{1/3}$ for bounded-error randomized
query complexity. In the first two theorems, $f$ and $g$ are nonconstant
total Boolean functions on $n$ and $m$ bits, respectively. We write
$D=\deg(f)$ and $E_f=Q_E(f)$.

\begin{theorem}[Exact evaluation of compositions, Theorem~\ref{thm:main}]
\label{thm:intro-composition}
The composition $f\circ g$ satisfies
\begin{equation}\label{eq:intro-subset}
 Q_E(f\circ g)\le6mD+12E_fQ_{1/(4n)}(g).
\end{equation}
\end{theorem}

The simulation of the outer algorithm uses approximate inner
operations and accounts for the second term. The first term pays for
the correction, which reads at most $D$ inner blocks in each branch.

\begin{theorem}[Degree-dependent error threshold, Theorem~\ref{thm:degree-error}]
\label{thm:intro-degree-error}
The composition $f\circ g$ satisfies
\begin{equation}\label{eq:intro-main}
 Q_E(f\circ g)=O\!\left(mD+E_fQ_{1/(4D)}(g)\right).
\end{equation}
\end{theorem}

The error threshold depends only on degree, which can be much smaller
than the number of outer variables. Standard error reduction
gives $Q_E(f\circ g)=O(mD+E_fQ(g)\log(4D))$. The proof interpolates the
inner query operations. The return amplitude is a polynomial of degree
at most $D$, so higher-order contributions cancel in the amplitude
used by the algorithm.

\begin{theorem}[Recursive NAE, Theorem~\ref{thm:nae-near-tight}]
\label{thm:intro-nae}
Let $F_k=\NAE_3^{\circ k}$, where $\NAE_3$ (not all equal on $3$ bits) is zero precisely on the
two constant strings. For $a=3/\sqrt2$,
\begin{equation}\label{eq:intro-nae}
 \Omega(a^k)\le Q_E(F_k)
 \le a^k\exp\!\left(O((\log(k+2))^2)\right).
\end{equation}
\end{theorem}

The exact complexity has the same exponential rate as the
bounded-error complexity. The upper bound improves the rate
$2^{11/8}=2.593\ldots$ of Ambainis's STOC 2013 algorithm to
$3/\sqrt2=2.121\ldots$. The lower bound follows from the known base
adversary value and its composition law
\cite{AmbExactSTOC,AmbExact,ReichardtSpalek,LMRSS}.

\begin{theorem}[Exact versus randomized queries under recursion,
Theorem~\ref{thm:recursive-randomized-upper}]
\label{thm:intro-recursive-randomized}
For every fixed nonconstant total gadget $h$,
\begin{equation}\label{eq:intro-recursive-randomized-upper}
 Q_E(h^{\circ k})=\widetilde O_h\!\left(R(h^{\circ k})^2\right),
 \qquad Q_E(h^{\circ k})\le R(h^{\circ k})^{2+o_h(1)}.
\end{equation}
\end{theorem}

The tilde hides a polylogarithmic factor with a gadget-dependent
exponent. Theorem~\ref{thm:recursive-randomized-upper} bounds this factor
and identifies conditions under which it can be omitted. Its proof
uses the following characterization of the exact quantum composition limit.

\begin{theorem}[Exact quantum composition limit,
Theorem~\ref{thm:recursive-growth-rate}]
\label{thm:intro-rate}
For every nonconstant total Boolean function $h$,
\begin{equation}\label{eq:intro-rate}
 \QElim(h):=\lim_{k\to\infty}Q_E(h^{\circ k})^{1/k}
       =\max\{\deg(h),\Adv(h)\}.
\end{equation}
\end{theorem}

We set $\QElim(h)=0$ for constant $h$. We follow the notation $M^{\lim}$ of Gilmer, Saks, and Srinivasan for
composition limits. Al-Dhalaan and Ben-David use $M^*$ for the same
operation~\cite{GilmerSaksSrinivasan2013,AlDhalaanBenDavid2026}.
Degree and the adversary bound compose exactly and lower bound
$Q_E$ up to constants. Our algorithm attains the larger of their
exponential rates. Its quantitative bounds, together with
$\deg(F)\le9R(F)^2$ and $\Adv(F)=O(R(F))$, yield
\eqref{eq:intro-recursive-randomized-upper}
\cite{Huang2019,Nisan,Reichardt,LMRSS}.

\begin{theorem}[Transfer principles, Theorems~\ref{thm:tr-general}
and~\ref{thm:tr-bs} and Corollary~\ref{cor:tr-four-way}]
\label{thm:intro-transfer}
Fix $\alpha>0$. Let $L$ be a nonnegative measure on total Boolean
functions that vanishes on constants and satisfies
$L(f^{\circ k})\ge L(f)^k$ for every $f$ and $k\ge1$. Then
\begin{equation}\label{eq:intro-transfer}
 \bigl[L(F)=O(Q_E(F)^\alpha)\text{ for all }F\bigr]
 \quad\Longleftrightarrow\quad
 \bigl[L(f)\le \QElim(f)^\alpha\text{ for all }f\bigr].
\end{equation}
For each $B\in\{s,\bs,\fbs\}$, the universal bound
$B(F)=O(Q_E(F)^\alpha)$ is equivalent to
$\fbs(f)\le\QElim(f)^\alpha$ for all $f$.
\end{theorem}

Taking $k$th roots along recursive powers removes the universal
constant and gives a pointwise criterion with coefficient one. In the
reverse direction, degree and the adversary bound give
$\QElim(f)\le2Q_E(f)$. The fractional block-sensitivity criterion uses
the normalization and composition results of Tal and
Kulkarni--Tal~\cite{Tal,KT}.

\begin{theorem}[Sensitivity separation, Theorem~\ref{thm:sep-three-halves}]
\label{thm:intro-sensitivity-separation}
Let $m_k=\lceil(9/8)^k\rceil$, $G_k=F_k\circ\OR_{m_k}$, and
$N_k=3^km_k$. Then
\begin{equation}\label{eq:intro-separation}
 s(G_k)=\bs(G_k)=\fbs(G_k)=N_k,\qquad
 Q_E(G_k)=N_k^{2/3+o(1)}.
\end{equation}
For each $L\in\{s,\bs,\fbs\}$, the quantitative separation is
\[
 L(G_k)\ge Q_E(G_k)^{3/2}
     \exp\!\left(-O((\log\log(Q_E(G_k)+3))^2)\right).
\]
\end{theorem}

The OR layer preserves full sensitivity and balances the degree and
adversary costs. The exponent $3/2$ improves on
$8\log_2(3)/11=1.1527\ldots$, obtained from Ambainis's algorithm for
recursive NAE~\cite{AmbExactSTOC,AmbExact}.

\begin{theorem}[Randomized versus exact quantum queries,
Theorem~\ref{thm:randomized-quadratic}]
\label{thm:intro-randomized-separation}
There is a family of total Boolean functions $W_n$ for which
$q_n=Q_E(W_n)$ grows unboundedly and
\begin{equation}\label{eq:intro-randomized-separation}
 R(W_n)\ge q_n^2\exp\!\left(
   -O\!\left(\sqrt{\log(q_n+2)\log\log(q_n+3)}\right)\right).
\end{equation}
For every $0<\eps<1$, there is a fixed total function $h=h_\eps$ whose
recursive powers have unbounded exact quantum query complexity and satisfy
\[
 R(h^{\circ k})\ge Q_E(h^{\circ k})^{2-\eps}
\]
for all sufficiently large $k$.
\end{theorem}

This improves the previous exponent $3/2$ from pointer functions.
A cheat-sheet construction gives small degree and bounded-error
quantum complexity, while sabotage complexity supplies a randomized
lower bound with a base that is preserved under recursion
\cite{PointerFunctions,ABKCheatSheets,BKSabotage}. The composition
limit in~\eqref{eq:intro-rate} converts the two small parameters into
an exact quantum upper bound along the powers of a fixed $h$.
An explicit choice of recursion depth as a function of the base
gadget size gives the quantitative bound
in~\eqref{eq:intro-randomized-separation}.

\begin{theorem}[Different inner functions, Proposition~\ref{prop:sc-heterogeneous}]
\label{thm:intro-heterogeneous}
Let $g_i$ be nonconstant total Boolean functions on disjoint blocks
of sizes $m_i$, and let $f$ be nonconstant and total. If
$0<\eps_i<1/2$ and $\sum_i\eps_i\le1/4$, then
\[
 Q_E(f(g_1,\ldots,g_n))
 \le6D_{\mathbf m}(f)+12Q_E(f)\max_iQ_{\eps_i}(g_i).
\]
Here $D_{\mathbf m}(f)$ is the maximum sum of block sizes over the
support of a monomial of the signed outer polynomial.
\end{theorem}

The weighted degree records the largest correction cost of a branch.
The separate error budgets accommodate inner problems of different
difficulty.

\begin{theorem}[Communication separations,
Theorems~\ref{thm:communication} and~\ref{thm:communication-quadratic}]
\label{thm:intro-communication}
Let $\mathcal H_h$ denote the coordinatewise AND lift of $h$.
Then $R_{1/3}^{\mathrm{cc}}(\mathcal H_{G_k})=\Theta(N_k)$ and
\[
 \Omega(N_k^{2/3})\le Q_E^{\mathrm{cc}}(\mathcal H_{G_k})
 \le N_k^{2/3}\exp\!\left(O((\log\log(N_k+2))^2)\right).
\]
There is also a family of total communication functions $\mathcal W_n$
for which $q_n=Q_E^{\mathrm{cc}}(\mathcal W_n)$ grows unboundedly and
\[
 R_{1/3}^{\mathrm{cc}}(\mathcal W_n)
 \ge q_n^2\exp\!\left(
    -O\!\left(\sqrt{\log(q_n+2)\log\log(q_n+3)}\right)\right).
\]
For every $0<\eps<1$, there are total communication functions
$\mathcal F_k$ of unbounded exact quantum communication complexity
such that, for all sufficiently large $k$,
\[
 R_{1/3}^{\mathrm{cc}}(\mathcal F_k)
 \ge Q_E^{\mathrm{cc}}(\mathcal F_k)^{2-\eps}.
\]
The randomized lower bounds hold with public randomness and unrestricted
interaction. The exact quantum upper bounds use no prior entanglement.
\end{theorem}

The AND lift preserves a disjointness restriction, and the log-rank
bound gives the matching quantum exponent. The nearly quadratic gap
uses the cheat-sheet family, an inner-product lift, and the standard
query simulation for exact protocols. It improves the previous
communication separation exponent $3/2$
\cite{BWComm,CFKMP2021,BCW98,ABBDGJKLS}.

\subsection{Proof strategy}
The first algorithm starts with the signed multilinear polynomial $H$
of the outer function. For a fixed composed input, let $z_i\in\{-1,1\}$
be the correct sign of the $i$th inner answer and let $b_i\in[-1,1]$ be
the output bias of its approximate algorithm. Multilinearity gives the
exact identity
\[
 H(z)=\sum_{S\subseteq[n]}(z-b)_S\,\partial_SH(b).
\]
Only sets of size at most $D$ contribute. This identity suggests the
algorithm, but it does not by itself give a query implementation: the
biases $b_i$ are unknown and the coefficients need not be positive.

We resolve these two issues separately. Each derivative is represented
as a bounded correlation under a product distribution generated by the
inner algorithms. Conjugating the outer query operations implements this
correlation without explicitly estimating any $b_i$. The remaining
residual factors are implemented by reading the selected blocks. Their
small size controls the normalization of a coherent sum over $S$.
The sum has the correct output sign and an input-independent magnitude,
so exact amplitude amplification~\cite{BHMT} produces a clean phase.

The degree-dependent error threshold requires a different expansion.
We interpolate each approximate inner reflection toward the exact
reflection, using a parameter $\lambda$. The complete circuit then has
return amplitude $H(b+\lambda(z-b))$. This scalar polynomial has degree
at most $D$, even though the circuit's operator expansion may have
nonzero terms of higher order. We implement its coefficients through
order $D$ and bound their total normalization. A constant number of
repetitions of the inner algorithm gives the accuracy needed for this
bound.

For recursion, the composition theorem is applied to an exact outer tree
and an approximate inner tree of unequal depths. Their depths are chosen
to balance the degree cost against the adversary cost. The three possible
relations between these parameters lead to the quantitative bounds in
Section~\ref{sec:fixed-gadget}. The transfer principles and separation
results then follow by comparing exponential rates.

\subsection{Relation to previous work}
Our proofs use the polynomial method~\cite{BBCMW}, the general adversary
characterization~\cite{Reichardt,LMRSS}, and exact amplitude amplification
\cite{BHMT} as established tools. The composition algorithm develops the
approximate-evaluation and exact-correction approach beyond the OR inner
function treated by Pednekar and Podder~\cite{PP}. The transfer arguments
use the composition framework of Tal~\cite{Tal} and the fractional
block-sensitivity results of Kulkarni and Tal~\cite{KT}.
For the randomized separation, the cheat-sheet construction
\cite{ABKCheatSheets} and sabotage lower bounds~\cite{BKSabotage} are
inputs to our recursive exact algorithm. Similarly, the communication
applications use existing simulation, rank, and lifting theorems
\cite{BCW98,BWComm,CFKMP2021}. These literature results are stated with
citations at the points where they are used; the proofs focus on the new
algorithms and their consequences.

\paragraph{Organization.}\label{par:organization}
Section~\ref{sec:prelim} reviews the query model and the relevant
complexity measures. Section~\ref{sec:algorithm} gives the composition
algorithm, and Section~\ref{sec:degree-error} improves its error
threshold to depend on degree. Section~\ref{sec:nae} applies the
algorithm to recursive NAE. Section~\ref{sec:fixed-gadget} analyzes
recursive powers of a fixed function and compares exact and randomized
queries. Section~\ref{sec:tr-transfer} derives the transfer principles.
Sections~\ref{sec:sep-separations} and~\ref{sec:randomized-separation}
give the sensitivity and randomized query separations, respectively.
Section~\ref{sec:heterogeneous} handles different inner functions,
Section~\ref{sec:communication} gives the communication applications,
and Section~\ref{sec:conclusion} discusses open problems.

\section{Preliminaries}
\label{sec:prelim}

All functions are total and Boolean unless stated otherwise. We write
$[n]=\{1,\ldots,n\}$ and use $x^{(1)},\ldots,x^{(n)}$ for disjoint
input blocks. For nonconstant functions $f:\{0,1\}^n\to\{0,1\}$ and
$g:\{0,1\}^m\to\{0,1\}$, their block composition is
\[
 (f\circ g)(x^{(1)},\ldots,x^{(n)})
 =f(g(x^{(1)}),\ldots,g(x^{(n)})).
\]
When a composition has constant blocks, we first substitute their values
into the outer function. Logarithms are
natural unless a base is displayed. Constants are absolute except in
statements about a fixed recursive gadget, where they may depend on the
gadget. The notation $\widetilde O$ and $\widetilde\Omega$ suppresses
polylogarithmic factors in the stated parameter; a subscript $h$ permits
the constants and logarithmic exponents to depend on a fixed $h$.

\subsection{Query complexity and coherent algorithms}
We use the standard quantum query model~\cite{BBCMW,AmbExact}. For an
input $x\in\{0,1\}^N$, a phase query acts as
\[
 O_x\ket{j,w}=(-1)^{x_j}\ket{j,w}\quad(j\in[N]),
 \qquad O_x\ket{0,w}=\ket{0,w}.
\]
The dummy coordinate $0$ permits controlled queries and padding at unit
query cost. Algorithms may use arbitrary input-independent unitaries and
auxiliary registers. Only calls to $O_x$ are charged. We defer all
measurements until the end.

We write $Q_\eps(f)$ for the minimum worst-case number of queries of a
quantum algorithm whose error is at most $\eps$ on every input, and put
$Q(f)=Q_{1/3}(f)$ and $\QE(f)=Q_0(f)$. We use $D(f)$ for deterministic
query complexity and $R(f)=R_{1/3}(f)$ for randomized query complexity,
with the number of queries maximized over inputs and random choices.
Standard coherent error reduction gives~\cite{BuhrmanDeWolf2002}
\begin{equation}
 Q_\eps(f)=O\!\left(Q(f)\log(1/\eps)\right)
 \qquad(0<\eps\le1/3).
 \label{eq:error-reduction}
\end{equation}

Our constructions require the outer algorithm to return its workspace
to the initial state. The standard compute--phase--uncompute construction
achieves this at a factor-two query cost. Specifically, if $A_x$
is an exact $T$-query algorithm for $f$ and $Z$ acts on its output qubit,
then
\begin{equation}
 A_x^\dagger Z A_x\ket0=(-1)^{f(x)}\ket0.
 \label{eq:clean-phase}
\end{equation}
This is a \emph{clean phase} algorithm of cost $2T$. Conversely, a
controlled clean phase computes the corresponding Boolean value exactly
with the same query cost. We retain this factor of two throughout.

\subsection{Encoding scalars as return amplitudes}
The exact correction is assembled from circuits that encode real numbers
in a single amplitude. These amplitudes are used coherently; the
algorithm never estimates them. The \emph{return amplitude} of a unitary $U_x$ is $\bra0U_x\ket0$,
where $\ket0$ includes every initial register. If a $t$-query algorithm
$A_x$ has signed output bias $b(x)=\mathbb E[(-1)^{A(x)}]$, then
\begin{equation}
 \bra0A_x^\dagger Z A_x\ket0=b(x).
 \label{eq:bias-return}
\end{equation}
Running two circuits on disjoint work registers multiplies their return
amplitudes. To form a weighted sum, we use a selector register whose
amplitudes depend only on prescribed weights. For $w_j\ge0$, $L=\sum_jw_j>0$, and a selector
$V\ket0=\sum_j\sqrt{w_j/L}\ket j$,
\begin{equation}
 \bra{0,0}(V^\dagger\otimes I)
 \left(\sum_j\ket j\bra j\otimes U_{j,x}\right)
 (V\otimes I)\ket{0,0}
 =\frac1L\sum_jw_j\bra0U_{j,x}\ket0.
 \label{eq:weighted-return}
\end{equation}
The weights in this identity must be independent of the input. The
branch circuits may still depend on the input through their oracle calls.
Padding all branches to a common query schedule makes the total cost the
maximum branch cost.

The last step of both composition algorithms is exact amplitude
amplification~\cite{BHMT}. We use its standard three-call specialization:
a unitary with return amplitude in $\{-1/2,1/2\}$ can be converted to
the corresponding clean phase. The input-independent magnitude is what
makes this final amplification exact.

\subsection{Degree and the general adversary bound}
The exact degree $\deg(f)$ is the degree of the unique real multilinear
polynomial representing $f$ on the Boolean cube. The polynomial
method~\cite{BBCMW} gives
\begin{equation}
 \QE(f)\ge\tfrac12\deg(f).
 \label{eq:degree-lower}
\end{equation}
For an outer function $f$ we use its signed representation
\[
 H(y)=(-1)^{f((\mathbf1-y)/2)}
 =\sum_{T\subseteq[n]}\widehat H(T)y_T,
 \qquad y_T=\prod_{i\in T}y_i.
\]
The same symbol denotes its multilinear extension to $\mathbb R^n$.
We write $\partial_S=\prod_{i\in S}\partial/\partial y_i$ and, in
composition statements, abbreviate
\[
 D=\deg(f),\qquad E_f=\QE(f),\qquad E_g=\QE(g).
\]
Thus $D$ without an argument denotes polynomial degree, whereas $D(f)$
denotes deterministic query complexity.

We use the following exact composition law for degree.
\begin{lemma}[{Tal~\cite[Lemma~3.1(1)]{Tal}}]
\label{lem:degree-composition}
For nonconstant total Boolean functions $f$ and $g$,
$\deg(f\circ g)=\deg(f)\deg(g)$.
\end{lemma}

Midrijanis's deterministic algorithm~\cite{Midrijanis} gives
\begin{equation}\label{eq:det-degree-upper}
 \QE(f)\le D(f)\le2\deg(f)^3.
\end{equation}
A certificate at an input is a set of coordinates whose values force the
output. We write $\cert(f)$ for the maximum, over inputs, of the minimum
certificate size; in particular $\cert(f)\le D(f)$.

The second lower-bound measure that governs our recursive algorithms is
the general adversary bound:
\begin{equation}
 \Adv(f)=\max_{\Gamma\ne0}
 \frac{\|\Gamma\|}{\max_j\|\Gamma\circ\Delta_j\|}.
 \label{eq:adversary-primal}
\end{equation}
Here $\Gamma$ is real symmetric, indexed by inputs, and satisfies
$\Gamma_{xy}=0$ when $f(x)=f(y)$; the matrix $\Delta_j$ indicates
$x_j\ne y_j$. The symbol $\circ$ denotes entrywise product, and the
norm is the operator norm. We use the standard characterization and
composition theorems~\cite{Reichardt,LMRSS}:
\begin{equation}
 Q(f)=\Theta(\Adv(f)),\qquad
 \Adv(f\circ g)=\Adv(f)\Adv(g).
 \label{eq:adversary-composition}
\end{equation}
Their constants are universal. We also use the standard bounds
$\Adv(f)\le N$ for $N$-bit functions and $\Adv(f)\le2\QE(f)$.
Consequently, for each fixed $h$,
$Q(h^{\circ k})=\Theta(\Adv(h)^k)$ with constants independent of $k$.

The polynomial and adversary methods thus provide complementary lower
bounds. Their exact composition laws will identify the candidate
exponential rate for a recursive function; the main algorithm will give
the matching upper rate.

\section{An exact quantum algorithm for composed functions}
\label{sec:algorithm}

An approximate computation of an inner block can save queries, but an
exact outer algorithm cannot simply tolerate the resulting error. The
construction in this section corrects that error coherently. Its cost has
two contributions: simulating the outer queries with approximate inner
algorithms, and reading a small number of inner blocks to implement the
correction.

Throughout this section, $f:\{0,1\}^n\to\{0,1\}$ and
$g:\{0,1\}^m\to\{0,1\}$ are nonconstant total Boolean functions. Write
\[
 D=\deg(f),\qquad E_f=\QE(f),\qquad
 \eps=\frac1{4n},\qquad t=Q_\eps(g).
\]

\begin{theorem}[Exact evaluation of composed functions]
\label{thm:main}
For all such $f$ and $g$,
\begin{equation}
 \QE(f\circ g)\le 6mD+12E_ft.
 \label{eq:main-explicit}
\end{equation}
Moreover, there is a clean phase algorithm for $f\circ g$ with this query
cost. The same construction applies to any specified $t_A$-query algorithm
for $g$ with error at most $1/(4n)$, with $t_A$ in place of $t$.
\end{theorem}

\subsection{The algorithm}

Let $z_i$ be the correct sign of the $i$th inner output, and let $b_i$ be
the output bias of an approximate inner algorithm. The desired answer is
$H(z)$, where $H$ is the signed polynomial of $f$. Multilinearity gives an
exact expansion of $H(z)$ about $b$. Each term is indexed by a set $S$ of
at most $D$ inner blocks and consists of two factors: a derivative
$\partial_SH(b)$, and a product of residuals $z_i-b_i$ over $i\in S$.

We implement these two factors on separate work registers. The derivative
is a correlation of the Boolean function $H$ with a parity, so its
absolute value is at most one. More importantly, it can be implemented by
replacing each query of a clean outer circuit with a suitable inner
operation; this costs $4E_ft$ queries, independently of $|S|$ and $n$.
The residuals are small, and reading the blocks in $S$ lets us implement
their normalized product with $2m|S|$ additional queries.

The algorithm then combines all sets $S$ in a coherent superposition.
The weights depend only on $|S|$ and $\eps$, and their sum is less than
two. Padding this sum to exactly two produces a circuit whose return
amplitude is $H(z)/2$. Thus its success probability is the known constant
$1/4$, while its amplitude retains the sign of the answer. A standard
exact amplification step converts this amplitude into the clean phase
$H(z)$.

The rest of the proof establishes the expansion, implements one branch,
and then combines the branches. All appearances of $z_i$ and $b_i$ in
the analysis below describe the input and the inner algorithm; the
construction does not first compute the vectors $z$ or $b$.

\subsection{An exact expansion with bounded coefficients}

Let $H$ be the signed multilinear polynomial of $f$:
\[
 H(y)=(-1)^{f((\mathbf1-y)/2)}
      =\sum_{T\subseteq[n]}\widehat H(T)y_T,
 \qquad y\in\{-1,1\}^n,
 \qquad y_T=\prod_{i\in T}y_i.
\]
We use the same symbol for its multilinear extension to $\mathbb R^n$.
Fix a coherent $t$-query algorithm $A$ for $g$ with error at most $\eps$.
For an input $x=(x^{(1)},\ldots,x^{(n)})$ to the composed function, put
\begin{equation}
 z_i=(-1)^{g(x^{(i)})},\qquad
 b_i=\mathbb E[(-1)^{A(x^{(i)})}],\qquad
 e_i=z_i-b_i.
 \label{eq:sign-bias-residual}
\end{equation}
Here the expectation refers to measuring the output of $A$.
If its error probability on block $i$ is $p_i$, then
$b_i=z_i(1-2p_i)$ and $e_i=2z_ip_i$. In particular,
\[
 b_i\in[-1,1],\qquad |e_i|\le2\eps.
\]
The small size of $e_i$ will control the weight of every correction term.

For $S\subseteq[n]$, write
$\partial_S=\prod_{i\in S}\partial/\partial y_i$ and
$e_S=\prod_{i\in S}e_i$, with
$\partial_\varnothing H=H$ and $e_\varnothing=1$.
Expanding each monomial of $H(b+e)$ yields
\begin{equation}
 H(z)=\sum_{\substack{S\subseteq[n]\\|S|\le D}}
 e_S\,\partial_S H(b).
 \label{eq:taylor}
\end{equation}
There is no remainder: $H$ has degree $D$. There are also no factorials,
since a variable occurs at most once in each monomial. The following
identity expresses every derivative in~\eqref{eq:taylor} as a bounded
correlation, which is the form needed by the quantum circuit.

\begin{lemma}[Derivative correlations]
\label{lem:correlation}
For $S\subseteq[n]$ and $b\in[-1,1]^n$, let $Y_1,\ldots,Y_n$ be independent
signs such that $Y_i$ is uniform for $i\in S$ and
$\mathbb E Y_i=b_i$ for $i\notin S$. Then
\begin{equation}
 \partial_SH(b)=\mathbb E\!\left[H(Y)\prod_{i\in S}Y_i\right].
 \label{eq:derivative-correlation}
\end{equation}
Consequently, $|\partial_SH(b)|\le1$.
\end{lemma}
\begin{proof}
It suffices to evaluate the correlation on a monomial $Y_T$. If
$S\nsubseteq T$, choose $i\in S\setminus T$. The uniform sign $Y_i$
occurs exactly once in $Y_TY_S$, so independence makes the expectation
zero. If $S\subseteq T$, each sign indexed by $S$ occurs twice and
cancels, leaving expectation $\prod_{i\in T\setminus S}b_i$.
By linearity,
\[
 \mathbb E[H(Y)Y_S]
 =\sum_{T\supseteq S}\widehat H(T)\prod_{i\in T\setminus S}b_i
 =\partial_SH(b).
\]
The integrand $H(Y)Y_S$ takes values in $\{-1,1\}$, proving the bound.
\end{proof}

\subsection{Implementing the derivative correlation}
\label{subsec:lazy}

The correlation lemma suggests preparing all $n$ inner outputs and then
running the outer circuit. That preparation would cost $nt$ queries.
The next lemma obtains the same return amplitude by conjugating each
outer query separately, so that only the number of queries in the outer
algorithm enters the bound.

\begin{lemma}
\label{lem:derivative-circuit}
For every $S\subseteq[n]$, there is a unitary $\mathcal G_{S,x}$ with return amplitude
$\partial_SH(b)$ that uses at most $4E_ft$ queries to $x$. The construction
can be controlled coherently by a register containing $S$.
\end{lemma}
\begin{proof}
\proofstep{A circuit with the desired amplitude.}
Allocate an answer qubit $q_i$ and an inner workspace $w_i$ for each
block. When $i\notin S$, let $B_i=A_i$ be the algorithm $A$ on block
$x^{(i)}$, acting on $(q_i,w_i)$. When $i\in S$, let $B_i$ be a Hadamard
gate on $q_i$ and the identity on $w_i$. Define
\[
 B_S=\bigotimes_{i=1}^n B_i,
 \qquad P_S=\prod_{i\in S}Z_{q_i}.
\]
If the answer qubits in $B_S\ket0$ were measured, the signs
$Y_i=(-1)^{q_i}$ would have exactly the product distribution of
Lemma~\ref{lem:correlation}.

Let $C_f(q)$ be an exact $E_f$-query algorithm for $f$, with its own
workspace. Its clean phase circuit is
\[
 \mathcal W_f(q)=C_f(q)^\dagger Z_{\mathrm{out}}C_f(q),
\]
where $Z_{\mathrm{out}}$ acts on the outer answer qubit. This circuit
uses $2E_f$ queries to the virtual input $q$. For every basis string $q$,
\[
 \mathcal W_f(q)\bigl(\ket q\otimes\ket0_{\mathrm{outer}}\bigr)
 =(-1)^{f(q)}\ket q\otimes\ket0_{\mathrm{outer}}.
\]
Because the outer workspace is restored to zero, the same statement
holds coherently when $q$ is in superposition or entangled with the
inner workspaces. The factor $P_S$ contributes the additional sign
$\prod_{i\in S}(-1)^{q_i}$. Consequently,
\begin{equation}
 \bra0 B_S^\dagger P_S\mathcal W_f(q)B_S\ket0
 =\mathbb E[H(Y)Y_S]
 =\partial_SH(b).
 \label{eq:derivative-circuit}
\end{equation}
The state $\ket0$ in this expression includes both inner and outer
workspaces.

\proofstep{Implementing the circuit query by query.}
Equation~\eqref{eq:derivative-circuit} is used to identify the amplitude;
we do not implement it by preparing $B_S\ket0$. Instead, factor
$P_S\mathcal W_f$ into its individual gates and insert
$B_SB_S^\dagger$ between consecutive factors. This expresses its
conjugate as a product of conjugated gates. Every input-independent
outer gate commutes with $B_S$ and is therefore unchanged.

It remains to implement the conjugate of a virtual query. If $J$ is the
outer query-index register, that query is
\[
 O_q=\ket0\bra0_J\otimes I
       +\sum_{i=1}^n\ket i\bra i_J\otimes Z_{q_i}.
\]
Since $B_S$ is a tensor product over the blocks, its conjugate is
\begin{equation}
 B_S^\dagger O_qB_S
 =\ket0\bra0_J\otimes I
  +\sum_{i=1}^n\ket i\bra i_J\otimes B_i^\dagger Z_{q_i}B_i.
 \label{eq:conjugated-query}
\end{equation}
Thus only the workspace selected by $J$ is affected. If $i\notin S$,
the selected operation is $A_i^\dagger Z_{q_i}A_i$, which uses $2t$
queries. If $i\in S$, it is $\mathsf HZ\mathsf H=X_{q_i}$, which is
query-free. These alternatives can be implemented coherently in one
$2t$-query schedule: the raw input query is addressed to the selected
block, and inactive branches query the dummy coordinate. The remaining
factor also has a query-free conjugate,
\[
 B_S^\dagger P_SB_S=\prod_{i\in S}X_{q_i}.
\]
Replacing all $2E_f$ virtual queries therefore costs at most $4E_ft$
queries to $x$.

The work registers persist between virtual queries. Their states need
not be zero when a later selected operation is applied: the identities
above hold as operator identities on the entire Hilbert space. This is
why the query-by-query implementation agrees exactly with the circuit
in~\eqref{eq:derivative-circuit}, despite never preparing all inner
outputs at once. Finally, the choice between the two operations in
\eqref{eq:conjugated-query}, as well as all free gates, can be controlled
on a register storing $S$. This proves the coherent version of the lemma.
\end{proof}

\subsection{Implementing the normalized residuals}

The derivative circuit supplies the first factor in each Taylor term.
For the second factor, it is enough to realize $e_i/(2\eps)$ as a return
amplitude. Reading block $i$ makes this possible with no further calls
to the approximate algorithm.

\begin{lemma}
\label{lem:residual-lookup}
For a selected inner block $u\in\{0,1\}^m$, there is a $2m$-query unitary
with return amplitude
\[
 \frac{(-1)^{g(u)}-\mathbb E[(-1)^{A(u)}]}{2\eps}.
\]
\end{lemma}
\begin{proof}
Read all $m$ bits of $u$ into a register using $m$ queries. For each
possible value of that register, define
\[
 r(u)=\frac{(-1)^{g(u)}-\mathbb E[(-1)^{A(u)}]}{2\eps}.
\]
The error guarantee for $A$ implies $r(u)\in[-1,1]$. Apply to a fresh
qubit the real rotation
\[
 R_u=\begin{pmatrix}
 r(u)&-\sqrt{1-r(u)^2}\\
 \sqrt{1-r(u)^2}&r(u)
 \end{pmatrix},
\]
controlled by the register holding $u$, and then uncompute that register
with another $m$ queries. The return amplitude on the fresh qubit is
$r(u)$, and the block register has returned to zero.

This controlled rotation is an allowed input-independent operation:
$g$ and $A$ are fixed, so the numbers $r(u)$ form a fixed finite table.
The algorithm does not estimate an error probability at run time. The
table may be expensive to describe or implement in a gate model, but
such input-independent computation is free in the query model used here.
\end{proof}

Now fix a set $S$ with $|S|\le D$. Apply the residual construction to
each block in $S$, with a fresh work register for each application, and
apply the derivative circuit on a separate register. For a fixed input
$x$, return amplitudes on these registers multiply. The resulting
branch unitary $V_{S,x}$ therefore has return amplitude
\begin{equation}
 a_S(x)=\partial_SH(b)\prod_{i\in S}\frac{e_i}{2\eps}
 \label{eq:branch-amplitude}
\end{equation}
and query cost at most
\begin{equation}
 B=2mD+4E_ft.
 \label{eq:branch-budget}
\end{equation}
All branches use the same schedule: $D$ residual slots, followed by the
derivative circuit. Unused residual slots make dummy queries. Thus
selecting the branches coherently costs the maximum branch budget $B$,
rather than the sum of their budgets.

\subsection{Combining the branches and obtaining an exact answer}

\begin{proof}[Proof of Theorem~\ref{thm:main}]
\proofstep{Coherent summation.}
To recover the Taylor expansion from~\eqref{eq:branch-amplitude}, give
branch $S$ the weight $w_S=(2\eps)^{|S|}$. These weights are
input-independent and nonnegative. Their total satisfies
\begin{equation}
 L=\sum_{j=0}^D\binom nj(2\eps)^j
 \le(1+2\eps)^n
 \le\exp(2n\eps)=e^{1/2}<2.
 \label{eq:lcu-normalization}
\end{equation}
Add a branch indexed by $\bot$, of weight $2-L$, whose unitary has zero
return amplitude; a Pauli $X$ on a fresh qubit suffices. Prepare the
selector by an input-independent unitary $P$ satisfying
\[
 P\ket0=\sum_{\substack{S\subseteq[n]\\|S|\le D}}
       \sqrt{\frac{w_S}{2}}\ket S
       +\sqrt{\frac{2-L}{2}}\ket\bot.
\]
Apply the selected branch unitary and then $P^\dagger$. By coherent
selection, the resulting unitary $U_x$ has return amplitude
\begin{align}
 \alpha_x:=\bra0U_x\ket0
 &=\frac12\sum_{|S|\le D}w_Sa_S(x)\notag\\
 &=\frac12\sum_{|S|\le D}e_S\partial_SH(b)
 =\frac{H(z)}2\in\{-1/2,1/2\}.
 \label{eq:half-sign-amplitude}
\end{align}
The second equality cancels the residual normalizations in
\eqref{eq:branch-amplitude}; the last equality is the exact Taylor
identity~\eqref{eq:taylor}. The whole circuit uses at most $B$ queries.

\proofstep{Exact amplification.}
The magnitude of $\alpha_x$ is now known and independent of $x$. We use
the standard three-call specialization of exact amplitude amplification
\cite{BHMT}, also used by Pednekar and Podder~\cite{PP}. Let
$\Pi=\ket0\bra0$ project onto the joint all-zero state and set
$R=2\Pi-I$. For
\[
 K_x=-U_xRU_x^\dagger RU_x,
\]
the amplification identity gives
\begin{equation}
 \bra0K_x\ket0=3\alpha_x-4\alpha_x|\alpha_x|^2
              =3\alpha_x-4\alpha_x^3,
 \label{eq:three-call}
\end{equation}
where the last equality uses that $\alpha_x$ is real. Substituting
$\alpha_x=H(z)/2$ gives return amplitude $H(z)$. Since $K_x$ is unitary
and this amplitude has absolute value one, no amplitude remains in any
orthogonal state. Hence
\[
 K_x\ket0=H(z)\ket0
          =(-1)^{(f\circ g)(x)}\ket0.
\]
This is the required clean phase. The two copies of $U_x$ and one copy
of $U_x^\dagger$ use at most $3B=6mD+12E_ft$ queries.

Finally, the standard controlled clean-phase construction from
Section~\ref{sec:prelim} converts $K_x$ into an exact decision algorithm
with the same query cost. This proves~\eqref{eq:main-explicit} and both
claims of the theorem.
\end{proof}

\section{An error threshold controlled by the outer degree}
\label{sec:degree-error}

Theorem~\ref{thm:main} requires an inner error of order $1/n$.
The dependence on $n$ comes from summing over subsets of outer
coordinates: the normalization contains
$\sum_{j\le D}\binom nj(2\eps)^j$.
This section replaces that expansion by one indexed by the
\emph{query positions} of an exact outer algorithm. Its clean phase
implementation has only $2\QE(f)$ such positions.

The central point is that the number of query positions and the degree
play different roles. The former controls the normalization of the
expansion; the latter controls how many corrections a branch must
perform. More precisely, we construct a formal interpolation whose
return amplitude is
\[
 H(b+\lambda(z-b)).
\]
This scalar polynomial has degree at most $D=\deg(f)$, even though
the interpolating query operators need not be polynomial in $\lambda$.
We may therefore discard all terms of total order greater than $D$
after taking the return amplitude. A uniform bound on the operator
coefficients makes the remaining terms implementable at constant
normalization. The resulting error threshold initially depends on
$\QE(f)$; the known polynomial relation between $\QE(f)$ and degree
then converts it to a threshold depending on $D$, at constant overhead.

\begin{theorem}[Composition with a degree-controlled error threshold]
\label{thm:degree-error}
Let $f:\{0,1\}^n\to\{0,1\}$ and
$g:\{0,1\}^m\to\{0,1\}$ be nonconstant total Boolean functions, and
write $D=\deg(f)$ and $E_f=\QE(f)$. Then
\begin{equation}
 \QE(f\circ g)
 \le 6mD+12E_f Q_{1/(1024E_f^2)}(g).
 \label{eq:op-query-error}
\end{equation}
In particular,
\begin{equation}
 \QE(f\circ g)
 =O\!\left(m\deg(f)+\QE(f)Q_{1/(4\deg(f))}(g)\right).
 \label{eq:op-degree-error}
\end{equation}
An explicit choice of the universal constants is
\begin{equation}
 \QE(f\circ g)\le6mD+804E_fQ_{1/(4D)}(g).
 \label{eq:op-explicit-degree-error}
\end{equation}
The constructions also give clean phase algorithms with these query
costs.
\end{theorem}

\subsection{Overview of the algorithm}

Fix an inner algorithm $A$ of query cost $t$ and error at most
$\eps=1/(1024E_f^2)$. We use its compute--phase--uncompute reflection
as an approximate inner query. The proof has four parts.

First, for every inner input $u$, we interpolate between this reflection
and one that acts on the initial state with the correct sign
$z_u=(-1)^{g(u)}$. The expectation of the interpolated reflection
changes linearly with the interpolation parameter. Second, we show
that the Taylor coefficients of these reflections admit
\emph{input-independent} norm bounds whose sum is $O(\sqrt\eps)$.
Third, we substitute the reflections into the clean outer circuit and
show that its return amplitude has degree at most $D$. Finally, we
implement the terms of Taylor order at most $D$ and combine them
coherently.

The interpolation is a device for specifying the correction operators;
the algorithm does not need to implement an interpolating reflection
at an arbitrary parameter. A zero-order coefficient is implemented
directly by $A^\dagger Z_{\mathrm{out}}A$. A positive-order coefficient
is implemented by reading the selected inner block and using the
resulting string to control a fixed unitary dilation. A retained term
contains at most $D$ positive-order factors, so these full-block reads
cost at most $2mD$. The remaining query cost is at most $4E_ft$.
The coefficient bounds make the total normalization less than two,
and exact amplitude amplification completes the construction.

\subsection{An interpolation of the inner reflection}

For an inner input $u$, let $A_u$ denote the unitary circuit of $A$,
including all its work registers. Define
\[
 z_u=(-1)^{g(u)},\qquad p_u=p_{\mathrm{err}}(u),\qquad
 V_u=A_u^\dagger Z_{\mathrm{out}}A_u.
\]
Thus $V_u$ is a Hermitian unitary of query cost $2t$, and
\begin{equation}
 \bra0V_u\ket0=z_u(1-2p_u)=:b_u.
 \label{eq:op-initial-bias}
\end{equation}
We construct a family of reflections for which this expectation moves
from the approximate sign $b_u$ to the exact sign $z_u$.

\begin{lemma}[Interpolating reflection]
\label{lem:op-reflection}
For every inner input $u$, there is an operator-valued function
$V_u(\lambda)$ with the following properties:
\begin{enumerate}
\item $V_u(0)=V_u$;
\item $V_u(\lambda)$ is a Hermitian unitary for real
$0\le\lambda\le1$;
\item if $e_u=z_u-b_u=2z_up_u$, then
\begin{equation}
 \bra0V_u(\lambda)\ket0=b_u+\lambda e_u;
 \label{eq:op-affine-bias}
\end{equation}
\item $V_u(1)\ket0=z_u\ket0$.
\end{enumerate}
Moreover, $V_u(\lambda)$ has a Taylor series about zero that converges
absolutely in operator norm for $|\lambda|\le1$.
\end{lemma}

\begin{proof}
Write $p=p_u$ and $z=z_u$. If $p=0$, then
$V_u\ket0=z\ket0$, and the constant family
$V_u(\lambda)=V_u$ has every required property. We may therefore assume
$0<p\le\eps<1/2$.

Only a two-dimensional subspace needs to change. Define
\[
 \ket{r_u}
 =\frac{zV_u\ket0-(1-2p)\ket0}{2\sqrt{p(1-p)}}.
\]
The numerator is orthogonal to $\ket0$ by
\eqref{eq:op-initial-bias}; its squared norm is $4p(1-p)$ because
$V_u$ is unitary. Thus $\ket0,\ket{r_u}$ is an orthonormal pair.
Their span, denoted $\mathcal P_u$, is invariant under $V_u$.
Indeed, the definition gives $V_u\ket0$ in this span, and applying
$V_u$ again determines $V_u\ket{r_u}$. In this basis,
\[
 V_u\big|_{\mathcal P_u}
 =z\begin{pmatrix}
       1-2p&2\sqrt{p(1-p)}\\
       2\sqrt{p(1-p)}&-1+2p
    \end{pmatrix}.
\]
Since $V_u$ is Hermitian, $\mathcal P_u^\perp$ is invariant as well.

We now reduce the parameter $p$ continuously to zero. On
$\mathcal P_u$, set
\begin{equation}
 V_u(\lambda)\big|_{\mathcal P_u}
 =z\begin{pmatrix}
  1-2p+2p\lambda&
  2\sqrt p\sqrt{(1-\lambda)(1-p+p\lambda)}\\
  2\sqrt p\sqrt{(1-\lambda)(1-p+p\lambda)}&
  -1+2p-2p\lambda
 \end{pmatrix},
 \label{eq:op-reflection-matrix}
\end{equation}
and leave the action on $\mathcal P_u^\perp$ unchanged.
For real $\lambda\in[0,1]$, the displayed matrix is
\[
 z\begin{pmatrix}
  1-2p'&2\sqrt{p'(1-p')}\\
  2\sqrt{p'(1-p')}&-1+2p'
 \end{pmatrix},
 \qquad p'=p(1-\lambda).
\]
Its square is the identity, so it remains a Hermitian reflection.
At $\lambda=0$ it equals the original matrix. Its first diagonal
entry is $b_u+2z_up_u\lambda$, which proves
\eqref{eq:op-affine-bias}. At $\lambda=1$, its off-diagonal entries
vanish and its action on $\ket0$ is multiplication by $z_u$.

For the convergence statement, factor the off-diagonal entry as
\begin{equation}
 2z\sqrt{p(1-p)}\sqrt{1-\lambda}\sqrt{1+q\lambda},
 \qquad q=\frac p{1-p}<1.
 \label{eq:op-sqrt-factorization}
\end{equation}
The binomial series of $\sqrt{1-\lambda}$ converges absolutely on
the closed unit disk. The series of $\sqrt{1+q\lambda}$ has radius
$1/q>1$. Their product therefore converges absolutely for
$|\lambda|\le1$. The diagonal entries are affine and the orthogonal
complement is constant, proving convergence in operator norm.
\end{proof}

\subsection{A uniform bound on the Taylor coefficients}

A bound on the coefficients for each individual input would not yet
suffice: the amplitudes used to prepare the branch selector must be
fixed before seeing the input. We therefore bound every coefficient
by a common sequence that depends only on the promised error $\eps$.

\begin{lemma}[Uniform coefficient majorant]
\label{lem:op-majorant}
Write
\begin{equation}
 V_u(\lambda)=V_u(0)+\sum_{k\ge1}\lambda^k C_{u,k}.
 \label{eq:op-series}
\end{equation}
There are nonnegative numbers $w_k$, depending only on $\eps$, such that
\begin{equation}
 \|C_{u,k}\|\le w_k\quad\text{for every }u,k,
 \qquad \sum_{k\ge1}w_k\le8\sqrt\eps.
 \label{eq:op-majorant}
\end{equation}
Each $C_{u,k}$ is Hermitian.
\end{lemma}

\begin{proof}
The only nonconstant entries occur in the two-dimensional matrix
\eqref{eq:op-reflection-matrix}. To bound its off-diagonal coefficients
uniformly over $p\le\eps$, put
\[
 q_*=\frac{\eps}{1-\eps},\qquad
 a_j=\left|\binom{1/2}{j}\right|\quad(j\ge0),\qquad
 \gamma_k=\sum_{j=0}^k a_j a_{k-j}q_*^{\,k-j}.
\]
Here $a_0=\gamma_0=1$. The coefficient of $\lambda^k$ in
$\sqrt{1-\lambda}\sqrt{1+(p/(1-p))\lambda}$ has absolute value at
most $\gamma_k$: apply the triangle inequality to the product of the
two binomial series and use $p/(1-p)\le q_*$.
After restoring the factor in
\eqref{eq:op-sqrt-factorization}, the off-diagonal coefficient has
operator norm at most $2\sqrt\eps\,\gamma_k$.

The diagonal part contributes only at order one, with norm $2p$.
Thus the following choice bounds the complete coefficient:
\begin{equation}
 w_k=2\eps\,\mathbf1_{\{k=1\}}
                  +2\sqrt\eps\,\gamma_k,
 \qquad k\ge1.
 \label{eq:op-explicit-majorant}
\end{equation}
For $p=0$ all positive-order coefficients vanish, so the same bound
applies. For $p>0$ they vanish on $\mathcal P_u^\perp$; hence there
is no additional contribution from the rest of the workspace.

It remains to sum these bounds. The absolute binomial coefficients
satisfy
\[
 \sum_{j\ge0}a_j=2,\qquad
 \sum_{j\ge0}a_jq_*^j=2-\sqrt{1-q_*}\le2.
\]
The second identity is valid because
$q_*<1$, as $\eps\le1/1024$.
The convolution defining $\gamma_k$ therefore gives
\[
 \sum_{k\ge1}\gamma_k
 =\left(\sum_{j\ge0}a_j\right)
   \left(\sum_{j\ge0}a_jq_*^j\right)-1
 =3-2\sqrt{1-q_*}\le3.
\]
Substituting into \eqref{eq:op-explicit-majorant}, we obtain
\[
 \sum_{k\ge1}w_k\le2\eps+6\sqrt\eps\le8\sqrt\eps.
\]
Finally, the matrix in \eqref{eq:op-reflection-matrix} has Hermitian
Taylor coefficients: each is a real linear combination of the fixed
diagonal and off-diagonal Hermitian matrices on $\mathcal P_u$.
This also proves the assertion on the full workspace.
\end{proof}

\subsection{The return amplitude has low degree}

Fix a composed input $x=(x^{(1)},\ldots,x^{(n)})$. We allocate
\emph{one persistent inner workspace per block}. The same workspace
is used whenever the outer circuit queries that block. Write
$V_i(\lambda)=V_{x^{(i)}}(\lambda)$ for the reflection on the $i$th
workspace. With $J$ denoting the outer query-index register and
zero its dummy index, the virtual oracle is
\begin{equation}
 O_x(\lambda)=\ket0\bra0_J\otimes I
       +\sum_{i=1}^n\ket i\bra i_J\otimes V_i(\lambda).
 \label{eq:op-virtual-oracle}
\end{equation}
Let $C_f$ be an exact $E_f$-query algorithm for $f$. Replace each of
the $T=2E_f$ queries in its clean phase circuit
$C_f^\dagger Z_{\mathrm{out}}C_f$ by $O_x(\lambda)$, and denote
the resulting circuit by $\mathcal U_x(\lambda)$.
The outer gates act only on the outer registers. For real
$\lambda\in[0,1]$, every substituted oracle is unitary.

\begin{lemma}[Polynomial return amplitude]
\label{lem:op-polynomial}
Let $H$ be the signed multilinear polynomial of $f$, and set
$b_i=z_i(1-2p_i)$ and $e_i=2z_ip_i$ for each inner block. Then
\begin{equation}
 P_x(\lambda):=\bra0\mathcal U_x(\lambda)\ket0
             =H(b+\lambda e).
 \label{eq:op-return-polynomial}
\end{equation}
In particular, $P_x$ has degree at most $D$ and
$P_x(1)=H(z)=(-1)^{(f\circ g)(x)}$.
\end{lemma}

\begin{proof}
Fix real $\lambda\in[0,1]$. The reflections
$V_1(\lambda),\ldots,V_n(\lambda)$ act on disjoint registers,
so they commute and admit a joint spectral decomposition.
Their joint eigenvalues are sign strings $y\in\{-1,1\}^n$.
Conditioned on such a sign string, \eqref{eq:op-virtual-oracle}
is precisely the usual outer phase oracle with input signs $y$.
The exact outer circuit consequently returns its workspace to zero
and contributes the clean phase $H(y)$.

To compute the return amplitude, it remains to average these phases
over the spectral distribution of the initial inner state. That state
is a product of all-zero states, so its spectral signs
$Y_1,\ldots,Y_n$ are independent. Their expectations are
\[
 \mathbb E Y_i
 =\bra0V_i(\lambda)\ket0=b_i+\lambda e_i.
\]
We therefore have
\[
 P_x(\lambda)
 =\mathbb E H(Y)
 =\sum_{S\subseteq[n]}\widehat H(S)
       \prod_{i\in S}(b_i+\lambda e_i)
 =H(b+\lambda e).
\]
The second equality uses independence, and the last uses
multilinearity of $H$. Every nonzero monomial has at most $D$
variables, proving the degree bound. At $\lambda=1$ the argument
of $H$ is $b+e=z$, giving the exact composed sign.
\end{proof}

For the coefficient comparison below, we also use this identity as
an identity of power series. Lemma~\ref{lem:op-reflection} makes
the circuit product analytic on $|\lambda|<1$. Its return amplitude
agrees there with the polynomial $H(b+\lambda e)$, since the two
analytic functions agree on the real interval $[0,1)$.
Their Taylor coefficients are therefore equal. Absolute convergence
on the closed unit disk permits evaluation at $\lambda=1$.

\subsection{Implementing the truncated operator expansion}

We now turn the preceding scalar identity into a finite quantum
circuit. Expanding a virtual query gives zero-order coefficient
$O_x(0)$ and, for $k\ge1$, coefficient
\[
 \mathcal C_{x,k}
 =\sum_{i=1}^n\ket i\bra i_J\otimes C_{x^{(i)},k}.
\]
The dummy-index block of $\mathcal C_{x,k}$ is zero. This is a
direct sum over the query index, not a sum acting on the same index
subspace; hence $\|\mathcal C_{x,k}\|\le w_k$.

\paragraph{Which terms are retained.}
A term in the expanded outer circuit is indexed by
$\boldsymbol k=(k_1,\ldots,k_T)\in\mathbb Z_{\ge0}^T$.
In query position $j$, it uses the coefficient of order $k_j$;
all input-independent outer gates retain their original positions.
The resulting factors are multiplied in circuit order. Their total
Taylor order is $|\boldsymbol k|=\sum_jk_j$.
Let $a_{\boldsymbol k}(x)$ be the return amplitude of this
unnormalized product.

Retain all terms with $|\boldsymbol k|\le D$. There are finitely many,
and
\begin{equation}
 \sum_{|\boldsymbol k|\le D}a_{\boldsymbol k}(x)
 =P_x(1)=H(z).
 \label{eq:op-exact-truncation}
\end{equation}
To justify this equality, first expand the whole circuit. The coefficient
majorant makes the expansion absolutely convergent, so we may group
terms by total order. The sum at each order is the corresponding Taylor
coefficient of $P_x$. Lemma~\ref{lem:op-polynomial} says that these
\emph{scalar} coefficients vanish above order $D$. Thus truncation
preserves the return amplitude at one. No claim is made that the
individual high-order operator products vanish.

\paragraph{How a coefficient is implemented.}
Set $w_0=1$. A zero-order factor is implemented with $2t$ queries
by applying $A$, the output phase, and $A^\dagger$ to the selected
block's workspace. That workspace need not still be zero:
$V_u=A_u^\dagger Z_{\mathrm{out}}A_u$ is an operator identity,
so this implementation is valid on every intermediate state.

For a positive-order factor, the normalized coefficient
$K_{u,k}=C_{u,k}/w_k$ is a Hermitian contraction. Fix its unitary
dilation
\[
 \mathcal D(K_{u,k})=
 \begin{pmatrix}
  K_{u,k}&-\sqrt{I-K_{u,k}^2}\\
  \sqrt{I-K_{u,k}^2}&K_{u,k}
 \end{pmatrix}.
\]
The displayed blocks refer to a new ancilla qubit. Its zero-to-zero
block is $K_{u,k}$.
To apply this dilation, coherently read the selected input block
into a register, apply the dilation controlled by the resulting
string $u$, and uncompute the register. This costs $2m$ queries.
For every possible $u$ and $k$, the matrix of the dilation is fixed
in advance as part of the circuit description. Thus its controlled
application is an input-independent unitary once the string $u$
has been queried. Arbitrary input-independent unitaries are free
in the query model. At the dummy index, use a query-free dilation
whose zero-to-zero block is the zero operator.

Each positive-order position receives a fresh dilation ancilla, and
no later operation acts on that ancilla. Consequently, taking the
zero-to-zero block on all these ancillas gives exactly the ordered
product of the normalized coefficients. This observation is important:
reusing a dilation ancilla could allow amplitude that had left its
zero state to return later, producing additional terms.
With fresh ancillas, the branch indexed by $\boldsymbol k$ has
return amplitude
\[
 \frac{a_{\boldsymbol k}(x)}{w_{\boldsymbol k}},
 \qquad
 w_{\boldsymbol k}=\prod_{j=1}^T w_{k_j}.
\]
A branch with zero weight contributes nothing and may be omitted.

\paragraph{Why the branch cost depends on degree.}
Every positive-order position contributes at least one to
$|\boldsymbol k|$. A retained branch therefore contains at most $D$
such positions. There are at most $T$ zero-order positions, so a
uniform upper bound on the branch cost is
\begin{equation}
 B=2mD+2tT=2mD+4E_ft.
 \label{eq:op-branch-cost}
\end{equation}
This bound also applies to a coherent choice of branch. To see this,
pad each \emph{complete branch} to $B$ raw input queries.
Between successive raw queries, apply the gates appropriate to that
branch, controlled by the branch selector, and route the next query
to its required input coordinate. Dummy queries supply the padding.
All branches then use the same raw-query schedule.
Padding each virtual-query position separately would unnecessarily
charge the larger of the two slot costs at every position; aligning
the complete branches preserves \eqref{eq:op-branch-cost}.

\begin{proof}[Proof of Theorem~\ref{thm:degree-error}]
Choose $A$ with error $\eps=1/(1024E_f^2)$ and query cost
$t=Q_\eps(g)$. We have implemented each retained term at the common
cost $B$ in \eqref{eq:op-branch-cost}. We first bound the total weight
needed to combine these terms:
\begin{align}
 L
 &=\sum_{|\boldsymbol k|\le D}
                         \prod_{j=1}^T w_{k_j}\notag\\
 &\le\left(1+\sum_{k\ge1}w_k\right)^T
 \le(1+8\sqrt\eps)^{2E_f}
 \le\exp(16E_f\sqrt\eps)
 =e^{1/2}<2.
 \label{eq:op-normalization}
\end{align}
The first inequality only adds nonnegative terms. The second is the
uniform majorant from Lemma~\ref{lem:op-majorant}. Crucially, $L$
and all the branch weights are independent of the composed input.

Prepare a selector with amplitude
$\sqrt{w_{\boldsymbol k}/2}$ on each retained branch.
Complete its normalization with a query-free branch of weight $2-L$
and return amplitude zero. Apply the selected branch circuit and
unprepare the selector. The resulting return amplitude is
\[
 \frac12\sum_{|\boldsymbol k|\le D}
       w_{\boldsymbol k}\frac{a_{\boldsymbol k}(x)}{w_{\boldsymbol k}}
 =\frac{H(z)}2,
\]
where the equality is \eqref{eq:op-exact-truncation}.
The three-call identity \eqref{eq:three-call}, a specialization of
exact amplitude amplification~\cite{BHMT}, turns this into return
amplitude $H(z)$. Since $|H(z)|=1$, the whole final state is
$H(z)\ket0$; all work registers are clean. The cost is
$3B=6mD+12E_ft$, proving \eqref{eq:op-query-error}.

We next replace the error parameter by one depending only on degree.
The known deterministic degree bound~\cite{Midrijanis} gives
\begin{equation}
 E_f\le D(f)\le2\deg(f)^3=2D^3,
 \label{eq:op-deterministic-degree}
\end{equation}
where $D(f)$ denotes deterministic query complexity.
It is therefore enough to reduce the inner error to
$1/(4096D^6)$. Start with an algorithm of error at most $1/(4D)$,
run $67$ independent copies, and take their majority.
For $D\ge2$, an incorrect majority requires at least $34$ failed
copies, so its probability is at most
\[
 2^{67}\left(\frac1{4D}\right)^{34}
 =\frac1{2D^{34}}
 \le\frac1{4096D^6}
 \le\frac1{1024E_f^2}.
\]
For $D=1$, the standard majority tail bound gives error at most
$\exp(-67/8)<1/4096$, which is again sufficient.
The repetitions and the majority computation can be performed
coherently, retaining the work registers of all copies. Their total
query cost is $67Q_{1/(4D)}(g)$.
Substituting this cost into \eqref{eq:op-query-error} proves
\eqref{eq:op-explicit-degree-error}, and hence
\eqref{eq:op-degree-error}.
\end{proof}

The correction cost in this proof is a full-block read, not merely an
exact evaluation of the inner output. The operator
$C_{u,k}/w_k$ depends on the error direction $\ket{r_u}$ and the error
probability $p_u$, as well as on $g(u)$. Knowing $g(u)$ alone does not
specify the dilation that the correction must apply. This is the
specific obstacle to replacing $m$ by $\QE(g)$ in the present
construction; we return to that question in
Section~\ref{sec:conclusion}.

\section{An improved exact algorithm for recursive NAE}
\label{sec:nae}
Let $\NAE_3:\{0,1\}^3\to\{0,1\}$ be the not-all-equal function,
which is zero precisely on $000$ and $111$. Define $F_0(x)=x$ and
$F_d=\NAE_3\circ F_{d-1}$ for $d\ge1$. Thus $F_d$ is computed by the
complete ternary tree of depth $d$ with a copy of $\NAE_3$ at each internal
vertex. Write
\[
 E_d=\QE(F_d),\qquad a=\frac{3}{\sqrt2}.
\]
Ambainis~\cite{AmbExactSTOC,AmbExact} gave an exact algorithm using
$O(2^{11d/8})=O(2.593\ldots^d)$ queries. We obtain an upper bound whose
exponential rate matches the general adversary lower bound. The useful
feature of this family is that its degree, bounded-error complexity,
and input length grow at three different exponential rates. We exploit
this difference by applying the composition theorem at a carefully
chosen level of the tree.

\begin{theorem}\label{thm:nae-near-tight}
For all $d\ge0$,
\begin{equation}\label{eq:nae-near-tight}
 \Omega(a^d)\le \QE(F_d)
 \le a^d\exp\!\left(O((\log(d+2))^2)\right).
\end{equation}
Consequently, for $N=3^d$,
\[
 \QE(F_d)=N^{\log_3(3/\sqrt2)+o(1)},
 \qquad \log_3(3/\sqrt2)=0.684535\ldots.
\]
\end{theorem}

\subsection{Degree and adversary bounds}
The construction uses the following standard parameters of recursive NAE.

\begin{lemma}\label{lem:nae-parameters}
For every $d\ge0$,
\[
 \deg(F_d)=2^d,\qquad \Adv(F_d)=a^d,\qquad Q(F_d)=\Theta(a^d),
\]
where the constants in the last comparison are independent of $d$.
\end{lemma}
The degree-two polynomial of $\NAE_3$ and the value
$\Adv(\NAE_3)=3/\sqrt2$ are known; the latter is listed for the
output-complementary function $\mathrm{EQUAL}_3$ in
Reichardt and \v{S}palek~\cite[Table~1]{ReichardtSpalek}.
The displayed identities follow from the standard degree and adversary
composition theorems and the characterization $Q=\Theta(\Adv)$
\cite{Tal,Reichardt,LMRSS}.

\subsection{The recursive construction}
Consider a split $d=k+\ell$. The top $k$ levels form an outer tree
$F_k$, and each of its leaves is replaced by an inner tree $F_\ell$.
Thus $F_d=F_k\circ F_\ell$. We construct the exact algorithm for $F_k$
recursively. For $F_\ell$, we use the bounded-error algorithm from
Lemma~\ref{lem:nae-parameters}, followed by standard coherent error
reduction. The accuracy required by Theorem~\ref{thm:degree-error} is
\[
 \eps_k=\frac1{4\deg(F_k)}=\frac1{4\cdot2^k}.
\]
Since $\log(1/\eps_k)=O(k+1)$, this inner algorithm uses
$O((k+1)a^\ell)$ queries. Substituting the outer degree $2^k$, the
inner input length $3^\ell$, and this query cost into
Theorem~\ref{thm:degree-error} gives, for $1\le k<d$,
\begin{equation}\label{eq:nae-recurrence}
 E_d\le C\left(2^k3^{d-k}+kE_k a^{d-k}\right),
\end{equation}
where $C$ is an absolute constant. The first term pays for reading the
inner blocks used in the exact correction. The second term pays for
substituting the approximate inner computations into the exact outer
algorithm.

Our target is a cost close to $a^d$. The second term already has this
exponential scale if $E_k$ has scale $a^k$. We therefore choose the split
so that the correction term is no larger than $a^d$. Solving
$2^k3^{d-k}=a^d$ for $k$ suggests
\begin{equation}\label{eq:nae-split}
 \beta=\frac{\log(3/a)}{\log(3/2)}=0.854755\ldots,
 \qquad k=\lceil\beta d\rceil,\qquad \ell=d-k.
\end{equation}
Rounding $k$ upward only decreases the correction term, since $2<3$.
Choose a fixed base depth $d_0$ large enough that $1\le k<d$ for
$d>d_0$. At depths at most $d_0$, the algorithm reads all leaves.
At a larger depth, it recursively constructs the exact algorithm for
$F_k$ and combines it with the above approximate algorithm for
$F_\ell$. This specifies an exact algorithm at every depth: the exact
recursive call has smaller depth, and every correction branch reads
at most $2^k$ inner blocks of $3^\ell$ bits each.

\begin{proof}[Proof of Theorem~\ref{thm:nae-near-tight}]
The choice of $k$ in~\eqref{eq:nae-split} gives
\[
 \frac{2^k3^{d-k}}{a^d}
 =\exp\!\left(d\log(3/a)-k\log(3/2)\right)\le1.
\]
It remains to control the overhead accumulated by the recursive exact
calls. Normalize by the desired exponential factor and write
$u_d=E_d/a^d$. Equation~\eqref{eq:nae-recurrence} becomes
\begin{equation}\label{eq:nae-normalized-recurrence}
 u_d\le C(1+ku_k).
\end{equation}
Let $V_d=\max\{1,u_d\}$. Increasing a constant $C_1$ if necessary, we have
\[
 V_d\le C_1(d+1)V_{\lceil\beta d\rceil}
 \qquad(d>d_0).
\]
Thus each recursive call incurs a factor $O(d)$ in the normalized cost,
while reducing the depth by a constant factor. To make this precise, put
$d_0'=d$ and $d_{j+1}'=\lceil\beta d_j'\rceil$, stopping when
$d_j'\le d_0$. Enlarge the fixed base threshold so that
$d_0>2/(1-\beta)$. The accumulated rounding errors satisfy
\[
 d_j'\le\beta^j d+\frac1{1-\beta}.
\]
Hence there are $L=O(\log(d+2))$ stages. Multiplying the factors along
this sequence, and using $d_j'\le d$ before termination, gives
\[
 V_d\le\left(\max_{0\le u\le d_0}V_u\right)
          \prod_{j=0}^{L-1}C_1(d_j'+1)
 \le\exp\!\left(O((\log(d+2))^2)\right).
\]
Multiplication by $a^d$ proves the upper bound. The lower bound follows
from Lemma~\ref{lem:nae-parameters}, since
$\QE(F_d)\ge Q(F_d)=\Omega(a^d)$.
\end{proof}

Ambainis's construction~\cite{AmbExact} composes algorithms having a
fixed return amplitude on each output class and uses exact amplitude
amplification to improve their composition parameters. Here the
approximate inner algorithm may have an input-dependent error
probability. The correction in Theorem~\ref{thm:degree-error} removes
this error exactly. Choosing the split in~\eqref{eq:nae-split} makes
that correction affordable without changing the exponential rate
of the bounded-error algorithm.

\section{Exact algorithms for fixed recursive functions}
\label{sec:fixed-gadget}

The same construction applies to every family obtained by recursively
composing a fixed total Boolean gadget. We first state the resulting
comparison with randomized query complexity. The proof will identify
the exact exponential growth rate of the quantum algorithm, and then
use known randomized lower bounds to control that rate and its overhead.

\begin{theorem}[Exact quantum upper bounds in terms of randomized complexity]
\label{thm:recursive-randomized-upper}
Fix a total Boolean function $h:\{0,1\}^r\to\{0,1\}$ that depends
on at least two variables. Set
\[
 F_k=h^{\circ k},\qquad \delta=\deg(h),\qquad A=\Adv(h),
\]
and define
\begin{equation}\label{eq:recursive-randomized-log-exponent}
 p_h=
 \begin{cases}
 \displaystyle\frac{\log(r/\delta)}{\log(\delta/A)},&\delta>A,\\[6pt]
 0,&\delta\le A.
 \end{cases}
\end{equation}
Then there is a constant $C_h>0$, independent of $k$, such that
\begin{equation}\label{eq:recursive-randomized-polynomial-depth}
 \QE(F_k)\le C_h R(F_k)^2(k+1)^{p_h}
 \qquad(k\ge1).
\end{equation}
In particular,
\begin{equation}\label{eq:recursive-randomized-polylog}
 \QE(F_k)
 =O_h\!\left(R(F_k)^2
       [\log(R(F_k)+2)]^{p_h}\right),
 \qquad
 \QE(F_k)\le R(F_k)^{2+o_h(1)},
\end{equation}
where the last bound is as $k\to\infty$. If $\delta<A^2$ or
$\delta=r$, the stronger bound
$\QE(F_k)=O_h(R(F_k)^2)$ holds without a logarithmic factor.
\end{theorem}

The gadget is fixed throughout this statement. In particular, the
constant $C_h$ and the logarithmic exponent $p_h$ may depend on $h$.
The result does not assert a universal quadratic bound for individual
Boolean functions.

\subsection{The exact recursive rate}
Let $h:\{0,1\}^r\to\{0,1\}$ be a fixed nonconstant Boolean function,
and let $H_0(x)=x$ and $H_d=h^{\circ d}$. Write
\[
 T_d=\QE(H_d),\qquad \delta=\deg(h),\qquad A=\Adv(h),
 \qquad M=\max\{\delta,A\}.
\]
Degree composition forces a lower bound of order $\delta^d$, and
adversary composition forces one of order $A^d$. The next theorem shows
that the larger of these two bases also suffices for an exact algorithm,
up to a subexponential factor in the depth. All implicit constants in
its quantitative bounds may depend on $h$.

\begin{theorem}\label{thm:recursive-growth-rate}
For every fixed nonconstant Boolean function $h$,
\begin{equation}\label{eq:recursive-growth-rate}
 \QElim(h):=\lim_{d\to\infty}\QE(h^{\circ d})^{1/d}
 =\max\{\deg(h),\Adv(h)\}.
\end{equation}
More precisely, the following bounds hold:
\begin{enumerate}
\item If $\delta<A<r$, then
\[
 \Omega(A^d)\le T_d\le A^d\exp\!\left(O((\log(d+2))^2)\right).
\]
\item If $A<\delta$, then
\[
 \frac12\delta^d\le T_d\le O\!\left(\delta^d(d+1)^p\right),
 \qquad p=\frac{\log(r/\delta)}{\log(\delta/A)}.
\]
\item If $\delta=A<r$, then
\[
 \frac12\delta^d\le T_d
 \le\delta^d\exp\!\left(O(\sqrt{d\log(d+2)})\right).
\]
\item If $M=r$, then $T_d=\Theta(r^d)$.
\end{enumerate}
\end{theorem}

\begin{proof}
We derive one recurrence and choose its depth split differently in the
three nontrivial parameter regimes.

\proofstep{Lower bounds and the common recurrence.}
The standard composition theorems give
$\deg(H_d)=\delta^d$ and $\Adv(H_d)=A^d$. The polynomial method
and adversary characterization therefore imply
\begin{equation}\label{eq:fixed-gadget-lower}
 T_d\ge\tfrac12\delta^d,\qquad T_d=\Omega(A^d),
 \qquad T_d=\Omega(M^d).
\end{equation}
For the upper bound, write $d=k+\ell$, with $1\le k<d$, and view
$H_d$ as $H_k\circ H_\ell$. Its outer degree is $\delta^k$;
each inner block contains $r^\ell$ input bits and has bounded-error
query complexity $\Theta(A^\ell)$. Standard error reduction to
$1/(4\delta^k)$ uses $O_h(kA^\ell)$ queries. Thus
Theorem~\ref{thm:degree-error} gives
\begin{equation}\label{eq:general-recursive-recurrence}
 T_d\le C\left(\delta^k r^{d-k}+kT_kA^{d-k}\right),
\end{equation}
for a constant $C\ge1$ depending only on $h$. The first term is the
exact correction cost; the second uses the recursively constructed
exact outer algorithm. At any fixed set of initial depths, we may
instead read all input bits.

\proofstep{Adversary-dominated growth: $\delta<A<r$.}
Here the desired exponential base is $A$. As in the NAE construction,
we choose the split so that the correction term is at most $A^d$:
\[
 \beta_h=\frac{\log(r/A)}{\log(r/\delta)}\in(0,1),
 \qquad k=\lceil\beta_h d\rceil.
\]
For all sufficiently large $d$, this is a valid split, and
\[
 \frac{\delta^k r^{d-k}}{A^d}
 =\exp\!\left(d\log(r/A)-k\log(r/\delta)\right)\le1.
\]
Writing $u_d=T_d/A^d$, we obtain
$u_d\le C(1+ku_k)$. This is precisely the normalized recurrence
\eqref{eq:nae-normalized-recurrence}, with $\beta_h$ in place of
$\beta$. The exact recursive depth decreases by a constant factor,
so there are $O_h(\log(d+2))$ stages, each contributing at most a
factor $O_h(d+1)$. The product of these factors is
$\exp(O_h((\log(d+2))^2))$, proving the first upper bound.

\proofstep{Degree-dominated growth: $A<\delta$.}
If $\delta=r$, reading all input bits gives $T_d\le\delta^d$,
as required with $p=0$. Assume henceforth that $\delta<r$.
Normalize by the degree term, writing $u_d=T_d/\delta^d$.
With $\ell=d-k$, the recurrence becomes
\[
 u_d\le C(r/\delta)^\ell
       +C(d-\ell)(A/\delta)^\ell u_{d-\ell}.
\]
The two terms now favor opposite choices of $\ell$. A large inner
depth makes the correction expensive, but it decreases the coefficient
of the recursive term because $A/\delta<1$. We choose the smallest
order of depth that makes this coefficient at most $1/2$:
\[
 \ell=\left\lceil\frac{\log(2Cd)}{\log(\delta/A)}\right\rceil,
 \qquad k=d-\ell.
\]
For sufficiently large $d$, this gives $1\le k<d$. Moreover,
\[
 C(d-\ell)(A/\delta)^\ell\le\tfrac12,
 \qquad
 C(r/\delta)^\ell\le C_2(d+1)^p,
 \qquad
 p=\frac{\log(r/\delta)}{\log(\delta/A)},
\]
where $C_2$ absorbs the ceiling and depends only on $h$. Therefore
\[
 u_d\le C_2(d+1)^p+\tfrac12u_{d-\ell}.
\]
This inequality closes a polynomial induction. Choose $K\ge2C_2$
large enough that $u_j\le K(j+1)^p$ at every initial depth. Since
$p\ge0$, the induction hypothesis gives
\[
 u_d\le C_2(d+1)^p+\tfrac12K(d-\ell+1)^p
      \le K(d+1)^p.
\]
Multiplying by $\delta^d$ proves the second bound.

\proofstep{Equal growth rates: $\delta=A<r$.}
The preceding contraction is unavailable when $A=\delta$.
Set $u_d=T_d/\delta^d$ and $\lambda=\log(r/\delta)>0$.
Equation~\eqref{eq:general-recursive-recurrence} now gives
\begin{equation}\label{eq:critical-recursion}
 u_d\le C\left(e^{\lambda\ell}+d\,u_{d-\ell}\right).
\end{equation}
We allow a larger overhead and prove that
$u_d\le\exp(O_h(\sqrt{d\log(d+2)}))$. The split must make the
decrease in this proposed bound large enough to absorb the factor $d$,
while keeping the correction term within the same bound.

For real $u\ge2$, put $\Phi(u)=\sqrt{u\log u}$ and choose
$\ell=\lceil\Phi(d)\rceil$. At sufficiently large depths,
$\ell\le d/2$, and throughout $[d-\ell,d]$ we have
\[
 \Phi'(u)=\frac{\log u+1}{2\sqrt{u\log u}}
 \ge\frac13\sqrt{\frac{\log d}{d}}.
\]
Integrating this bound over an interval of length $\ell$ yields
\begin{equation}\label{eq:critical-potential-drop}
 \Phi(d)-\Phi(d-\ell)\ge\tfrac13\log d.
\end{equation}
This logarithmic decrease is what pays for the factor $d$ in the
recursive term.

Fix $K>\max\{\lambda,6\}$. We will prove
$u_d\le C_3e^{K\Phi(d)}$ by induction, where $C_3\ge1$ covers the
finitely many small depths. Assuming the bound at depth $d-\ell$,
divide the two terms in~\eqref{eq:critical-recursion} by
$C_3e^{K\Phi(d)}$. The correction term is at most
\[
 Ce^\lambda e^{-(K-\lambda)\Phi(d)},
\]
because $\ell\le\Phi(d)+1$. The recursive term is at most
\[
 Cd\,e^{-K(\Phi(d)-\Phi(d-\ell))}
 \le Cd^{1-K/3}.
\]
Both expressions tend to zero. Choose a fixed threshold beyond which
each is at most $1/2$, and then choose $C_3$ to cover all smaller
depths. The sum is at most one, completing the induction and the third
upper bound.

\proofstep{Full input-length growth: $M=r$.}
Reading the $r^d$ input bits gives $T_d\le r^d$, and
\eqref{eq:fixed-gadget-lower} supplies the matching lower bound.

In every parameter regime, the resulting bounds have the form
$\Omega(M^d)\le T_d\le M^d\exp(o_h(d))$. Taking $d$th roots
establishes both existence of the limit and the identity
\eqref{eq:recursive-growth-rate}.
\end{proof}

\subsection{Comparison with randomized query complexity}
\label{subsec:recursive-randomized-upper}
We now compare the algorithm with randomized query complexity.
Only the following known lower bounds on randomized algorithms are
needed; no composition theorem for $R$ is required.

\begin{lemma}\label{lem:randomized-controls-rate}
There is a universal constant $c_0>0$ such that every total Boolean
function $F$ satisfies
\begin{equation}\label{eq:randomized-controls-degree-adversary}
 \deg(F)\le9R(F)^2,
 \qquad
 \Adv(F)\le c_0R(F).
\end{equation}
Consequently, for nonconstant $F$,
$\QElim(F)\le\max\{9,c_0\}R(F)^2$.
\end{lemma}
The degree bound is the standard consequence of Huang's inequality
$\deg(F)\le s(F)^2$~\cite{Huang2019} and the randomized
block-sensitivity lower bound $\bs(F)\le3R(F)$~\cite{Nisan}.
The adversary bound follows from its characterization of bounded-error
quantum complexity and $Q(F)\le R(F)$~\cite{Reichardt,LMRSS}.
The final assertion uses Theorem~\ref{thm:recursive-growth-rate}.

\begin{proof}[Proof of Theorem~\ref{thm:recursive-randomized-upper}]
Write $r_k=R(F_k)$. Apply Lemma~\ref{lem:randomized-controls-rate}
to the entire depth-$k$ function, whose degree and adversary bound
compose exactly. This gives
\begin{equation}\label{eq:recursive-randomized-lower-ingredients}
 \delta^k=\deg(F_k)\le9r_k^2,
 \qquad
 A^k=\Adv(F_k)\le c_0r_k.
\end{equation}
The first inequality controls the degree contribution to our algorithm;
the second controls its adversary contribution.

Suppose first that $\delta>A$. By the second case of
Theorem~\ref{thm:recursive-growth-rate},
\[
 \QE(F_k)\le K_h\delta^k(k+1)^{p_h}
             \le9K_h r_k^2(k+1)^{p_h}.
\]
This proves~\eqref{eq:recursive-randomized-polynomial-depth} in the
degree-dominated case. If $\delta=r$, then $p_h=0$, so the same
argument already gives a bound without a logarithmic loss.

Now suppose that $\delta\le A$. The rate theorem gives
\[
 \QE(F_k)\le A^k\exp(o_h(k)).
\]
A total Boolean function of degree at most one depends on at most one
variable. Our hypothesis therefore implies $\delta\ge2$, and in this
case $A\ge2$. The extra factor $A^k$ available in the target quadratic
bound absorbs the entire subexponential overhead: there is a constant
$K_h$ such that $\exp(o_h(k))\le K_hA^k$ for every $k\ge1$.
Using the second inequality
in~\eqref{eq:recursive-randomized-lower-ingredients}, we obtain
\[
 \QE(F_k)\le K_hA^{2k}\le K_hc_0^2r_k^2.
\]
This includes the boundary case $\delta=A$: even its overhead
$\exp(O_h(\sqrt{k\log(k+2)}))$ is smaller than $A^k$ for all
sufficiently large $k$.

For the additional conclusion when $A<\delta<A^2$, use the same
slack in the adversary lower bound. Here
\[
 \delta^k(k+1)^{p_h}
 =A^{2k}(\delta/A^2)^k(k+1)^{p_h}
 =O_h(A^{2k}),
\]
since the exponentially decreasing factor $(\delta/A^2)^k$
absorbs the polynomial in $k$. Thus
$\QE(F_k)=O_h(r_k^2)$ also holds in this range.

Finally, we convert the polynomial overhead in the depth into a
polylogarithmic overhead in randomized complexity. Since $\delta\ge2$,
the first inequality in~\eqref{eq:recursive-randomized-lower-ingredients}
implies
\[
 r_k\ge\delta^{k/2}/3,
 \qquad k+1=O_h(\log(r_k+2)).
\]
Substitution proves~\eqref{eq:recursive-randomized-polylog}. The exponent
$p_h$ is fixed, while $r_k$ tends to infinity, so the logarithmic factor
and the fixed multiplicative constant are $r_k^{o_h(1)}$.
\end{proof}

If $h$ is constant, its recursive powers are constant. If it depends
on exactly one variable, every recursive power is a projection or a
negated projection, and both query complexities are one. These cases
satisfy the quadratic comparison directly.

\begin{remark}\label{rem:recursive-randomized-scope}
Theorem~\ref{thm:recursive-randomized-upper} applies to the powers of a
fixed gadget $h$. Its constants and logarithmic exponent may depend on $h$.
\end{remark}

\section{Transfer principles for exact-query lower bounds}
\label{sec:tr-transfer}

The preceding algorithm identifies the exponential growth of exact query
complexity under repeated composition. We now use that growth rate to
characterize inequalities that hold for \emph{all} total Boolean
functions. The principle is simple: apply a proposed universal inequality
to successive powers of one fixed function, and then take roots. The
constant in the universal inequality disappears, leaving a numerical
condition on the original function.

Recall from Theorem~\ref{thm:recursive-growth-rate} that
\begin{equation}\label{eq:tr-rate}
 \QElim(f)=\lim_{k\to\infty}\QE(f^{\circ k})^{1/k}
       =\max\{\deg(f),\Adv(f)\}.
\end{equation}
For a constant function we set $\QElim(f)=0$. Two properties explain why
this parameter is useful. First, the polynomial and adversary lower
bounds~\cite{BBCMW,Reichardt,LMRSS} imply
\begin{equation}\label{eq:tr-rate-upper}
 \QElim(f)\le2\QE(f).
\end{equation}
Second, exact composition of degree and of the adversary bound gives
\begin{equation}\label{eq:tr-rate-powers}
 \QElim(f^{\circ k})=\QElim(f)^k.
\end{equation}
Thus $\QElim$ is bounded by a constant multiple of exact query complexity,
but, unlike exact query complexity itself, it is exactly multiplicative
under self-composition.

\begin{theorem}[A criterion for universal exact-query bounds]
\label{thm:tr-general}
Fix a real number $\alpha>0$ and a nonnegative complexity measure $L$
on total Boolean functions. Assume that $L$ vanishes on constant
functions and that
\begin{equation}\label{eq:tr-supermultiplicative}
 L(f^{\circ k})\ge L(f)^k
 \qquad\text{for every total Boolean function $f$ and every integer $k\ge1$}.
\end{equation}
Then the following assertions are equivalent.
\begin{enumerate}
\item There is a constant $C>0$, depending at most on $L$ and $\alpha$,
such that every total Boolean function $F$ satisfies
\begin{equation}\label{eq:tr-universal}
 L(F)\le C\QE(F)^\alpha.
\end{equation}
In particular, $C$ is independent of $F$ and its number of variables.
\item Every total Boolean function $f$ satisfies
\begin{equation}\label{eq:tr-pointwise}
 L(f)\le\QElim(f)^\alpha
       =\max\{\deg(f),\Adv(f)\}^\alpha.
\end{equation}
\end{enumerate}
Moreover, assertion~2 implies assertion~1 with $C=2^\alpha$.
\end{theorem}

The coefficient one in assertion~2 is essential. A strict violation by
one finite function excludes \emph{every} universal constant in
assertion~1. The theorem therefore converts a question about an
asymptotic exponent into a search for a finite witness.

\begin{proof}
Suppose first that assertion~1 holds, and fix a nonconstant function
$f$. Apply the universal inequality to $f^{\circ k}$, rather than to
$f$ itself. By the assumed growth of $L$,
\[
 L(f)^k\le L(f^{\circ k})\le C\QE(f^{\circ k})^\alpha.
\]
Taking $k$th roots gives
\[
 L(f)\le C^{1/k}\bigl(\QE(f^{\circ k})^{1/k}\bigr)^\alpha.
\]
Both $f$ and $C$ are fixed while $k$ tends to infinity. The first factor
on the right tends to one; the second tends to $\QElim(f)^\alpha$ by
Theorem~\ref{thm:recursive-growth-rate}. This proves assertion~2.

For the converse, no composition is needed. Apply assertion~2 directly
to $F$ and then use~\eqref{eq:tr-rate-upper}:
\[
 L(F)\le\QElim(F)^\alpha\le2^\alpha\QE(F)^\alpha.
\]
This is assertion~1 with the stated constant. Constant functions satisfy
both assertions because $L$ and $\QE$ vanish on them.
\end{proof}

A strict violation $L(f)>\QElim(f)^\alpha$ supplies a separating family
$F_k=f^{\circ k}$. In fact, for
$\eta=\tfrac12\log(L(f)/\QElim(f)^\alpha)>0$ and all sufficiently large $k$,
\begin{equation}\label{eq:tr-strict-gap}
 L(F_k)\ge e^{\eta k}\QE(F_k)^\alpha.
\end{equation}
This follows from $L(F_k)\ge L(f)^k$ and
$\QE(F_k)=\QElim(f)^k\exp(o_f(k))$.

\subsection{Fractional block sensitivity and normalization}
\label{subsec:tr-fractional}

For block sensitivity, the appropriate finite-function criterion uses
\emph{fractional} block sensitivity. We first recall the definitions and
the composition facts needed for this passage; the latter are known
results of Tal and Kulkarni--Tal~\cite{Tal,KT}.

A nonempty set $B$ of coordinates is sensitive at $x$ if
$f(x^B)\ne f(x)$, where $x^B$ is obtained by flipping the coordinates
in $B$. Sensitivity $s(f,x)$ counts sensitive singletons. Block
sensitivity $\bs(f,x)$ is the largest number of pairwise disjoint
sensitive blocks. Fractional block sensitivity replaces disjointness by
a packing constraint: sensitive blocks receive nonnegative weights,
and the total weight using any one coordinate is at most one. Thus
\begin{equation}\label{eq:tr-fbs-definition}
 \fbs(f,x)=
 \max\left\{\sum_{B\text{ sensitive at }x}w_B:
     w_B\ge0,\quad\sum_{B\ni i}w_B\le1\text{ for every }i\right\}.
\end{equation}
The function measures are the maxima of their inputwise versions. For
an $n$-bit function they satisfy
\begin{equation}\label{eq:tr-sensitivity-chain}
 s(f)\le\bs(f)\le\fbs(f)\le n.
\end{equation}
We will use this chain to identify all three measures whenever a function
is sensitive to every coordinate at one input.

To iterate a fractional packing at zero, the output at zero must also be
zero: this ensures that the outer copy is evaluated at the same input
at every level. The following is the iterated form of Tal's normalized
composition inequality~\cite[Claim~5.7(ii)]{Tal}; see also
Kulkarni--Tal~\cite[Lemma~2.6(4)]{KT}.

\begin{lemma}[Fractional composition at the zero input; Tal~\cite{Tal}]
\label{lem:tr-normalized}
Let $h:\{0,1\}^n\to\{0,1\}$ satisfy $h(0^n)=0$, and put
$q=\fbs(h,0^n)$. Then, for every integer $k\ge1$,
\[
 \fbs(h^{\circ k},0^{n^k})\ge q^k.
\]
\end{lemma}

\begin{theorem}[A fractional criterion for universal block-sensitivity bounds]
\label{thm:tr-bs}
Fix $\alpha>0$. A universal constant $C>0$ satisfying
$\bs(F)\le C\QE(F)^\alpha$ for every total Boolean function $F$
exists if and only if every total Boolean function $f$ satisfies
$\fbs(f)\le\QElim(f)^\alpha$. Equivalently,
\begin{equation}\label{eq:tr-bs-equivalence}
\begin{aligned}
 &\bs(F)=O\!\left(\QE(F)^\alpha\right)\quad\text{for every }F\\
 &\hspace{12mm}\Longleftrightarrow\\[-2mm]
 &\fbs(f)\le\max\{\deg(f),\Adv(f)\}^\alpha
      \quad\text{for every }f.
\end{aligned}
\end{equation}
\end{theorem}

\begin{proof}
Assume the universal bound on $\bs$. Fix a nonconstant $n$-bit function
$f$. We will normalize an input attaining $\fbs(f)$, compose at that
input, and then use the known comparison of fractional and ordinary
block sensitivity along recursive powers.

\emph{Normalize the maximizing input.}
Choose $x^\star$ such that $\fbs(f,x^\star)=\fbs(f)$, and define
\[
 h(z)=f(z\oplus x^\star)\oplus f(x^\star).
\]
Input complementation moves $x^\star$ to zero, and output complementation
makes the value there zero. These operations preserve degree, the
adversary bound, and the sensitive-block structure. Consequently,
\[
 h(0)=0,\qquad \fbs(h,0)=\fbs(f),\qquad \QElim(h)=\QElim(f).
\]
Lemma~\ref{lem:tr-normalized} gives
$\fbs(h^{\circ k})\ge\fbs(f)^k$.

\emph{Pass from fractional to ordinary block sensitivity.}
Tal~\cite[Claim~5.10]{Tal} proved that the two measures differ by at most
a constant along powers of a fixed function. In the form recorded by
Kulkarni--Tal~\cite[Claim~2.7]{KT}, this says
\begin{equation}\label{eq:tr-constant-gap}
 \fbs(h^{\circ k})\le c(n)\bs(h^{\circ k}),
\end{equation}
where $c(n)$ depends only on the arity $n$ of the base function and is
independent of $k$. Combining this known result with the assumed
universal bound yields
\[
 \fbs(f)^k
 \le c(n)\bs(h^{\circ k})
 \le c(n)C\QE(h^{\circ k})^\alpha.
\]
Now take $k$th roots and let $k$ tend to infinity. Since the base function
is fixed, $(c(n)C)^{1/k}$ tends to one, and the recursive-rate theorem
gives
$\fbs(f)\le\QElim(h)^\alpha=\QElim(f)^\alpha$.

Conversely, the pointwise fractional bound immediately implies
\[
 \bs(f)\le\fbs(f)\le\QElim(f)^\alpha
       \le2^\alpha\QE(f)^\alpha.
\]
Constant functions cause no exception. This is the normalization and
root argument of Kulkarni--Tal~\cite[Lemma~3.1]{KT}, applied here with
our exact recursive rate.
\end{proof}

\begin{corollary}[The same universal exponent for three sensitivity measures]
\label{cor:tr-four-way}
For each $\alpha>0$, the following four statements are equivalent:
\begin{enumerate}
\item $s(F)=O(\QE(F)^\alpha)$ for every total Boolean function $F$;
\item $\bs(F)=O(\QE(F)^\alpha)$ for every total Boolean function $F$;
\item $\fbs(F)=O(\QE(F)^\alpha)$ for every total Boolean function $F$;
\item $\fbs(f)\le\max\{\deg(f),\Adv(f)\}^\alpha$ for every total
Boolean function $f$.
\end{enumerate}
Each implicit constant is independent of the function and its arity.
\end{corollary}

\begin{proof}
Theorem~\ref{thm:tr-bs}, equation~\eqref{eq:tr-rate-upper}, and the chain
$s\le\bs\le\fbs$ give all implications except $1\Rightarrow2$.
For this remaining implication, suppose that
$s(H)\le C\QE(H)^\alpha$ holds for every total Boolean function $H$.

Fix $F$, and choose an input $x$ with disjoint sensitive blocks
$B_1,\ldots,B_t$, where $t=\bs(F)$. Identify each block with a new
variable, defining
\[
 H(y)=F\!\left(x^{\,\bigcup_{j:y_j=1}B_j}\right).
\]
Every one of the $t$ variables is sensitive at zero, so $s(H)=t$.
Moreover, each input bit of $F$ becomes a constant, a variable of $H$,
or its complement. The standard query simulation for such substitutions
therefore gives $\QE(H)\le\QE(F)$, also for queries in superposition.
Applying the assumed sensitivity bound to $H$ now yields
\[
 \bs(F)=s(H)\le C\QE(H)^\alpha\le C\QE(F)^\alpha.
\]
This proves $1\Rightarrow2$ and completes the equivalence.
\end{proof}

\section{A \texorpdfstring{$3/2$}{3/2} separation between sensitivity and exact quantum queries}
\label{sec:sep-separations}

We apply the transfer principle to construct total functions $G_k$ on
$N_k$ variables for which
\[
 s(G_k)=\bs(G_k)=\fbs(G_k)=N_k,
 \qquad \QE(G_k)=N_k^{2/3+o(1)}.
\]
Theorem~\ref{thm:sep-three-halves} quantifies the loss in the
exponent-$3/2$ lower bound. This improves the exponent
$8\log_2(3)/11=1.152700\ldots$ supplied by Ambainis's STOC~2013
algorithm for recursive NAE~\cite{AmbExactSTOC,AmbExact}. That algorithm
uses $O(2048^{k/8})=O(2^{11k/8})$ queries on $3^k$ variables, all
sensitive at zero.

There are two steps in the construction. Recursive NAE supplies full
sensitivity and a small exact-query rate, but its degree is smaller than
its adversary bound. An additional OR layer brings these two parameters
into balance: an OR on $m$ bits multiplies degree by $m$ and the
adversary bound by $\sqrt m$. Choosing $m$ at the balancing point
improves the separation exponent to $3/2$. We first use the resulting
functions as finite witnesses for the transfer theorem, and then give a
direct quantitative bound for the same family.

\subsection{Full sensitivity of recursive NAE}
\label{subsec:sep-nae}

Let $F_k=\NAE_3^{\circ k}$ and $a=3/\sqrt2$. Section~\ref{sec:nae}
gives
\begin{equation}\label{eq:sep-nae-query}
 \deg(F_k)=2^k,\qquad \Adv(F_k)=a^k,\qquad
 \Omega(a^k)\le\QE(F_k)
 \le a^k\exp\!\left(O((\log(k+2))^2)\right).
\end{equation}
Recursive NAE is fully sensitive at zero, a property already used by
Ambainis~\cite[Section~3.1]{AmbExact}. The standard sensitivity,
certificate, and decision-tree inequalities therefore give
\begin{align}
 s(F_k)=\bs(F_k)=\fbs(F_k)&=3^k,
 \label{eq:sep-nae-sensitive}\\
 \cert(F_k)=D(F_k)&=3^k.
 \label{eq:sep-nae-certificate}
\end{align}
Write $\mathcal L=\{s,\bs,\fbs,\cert,D\}$ and
$\rho_{\NAE}=\log 3/\log(3/\sqrt2)=1.4608454206\ldots$.
Since $3=a^{\rho_{\NAE}}$, \eqref{eq:sep-nae-query} gives
\begin{equation}\label{eq:sep-nae-exponent}
 \frac{\QE(F_k)^{\rho_{\NAE}}}
      {\exp(O((\log\log(\QE(F_k)+3))^2))}
 \le L(F_k)\le O(\QE(F_k)^{\rho_{\NAE}})
 \qquad(L\in\mathcal L).
\end{equation}
The lower bound implies $L(F_k)\ge\QE(F_k)^\alpha$ for each fixed
$\alpha<\rho_{\NAE}$ and sufficiently large $k$. For the sensitivity
measures, the three-bit witness
$\fbs(\NAE_3)=3>\QElim(\NAE_3)^\alpha$ also gives this conclusion
through Corollary~\ref{cor:tr-four-way}.
In fact, the same quantitative bound gives
\[
 \lim_{k\to\infty}\frac{L(F_k)}{\QE(F_k)^\alpha}=\infty
 \qquad(\alpha<\rho_{\NAE},\ L\in\mathcal L),
\]
since $\QE(F_k)$ tends to infinity and the loss in
\eqref{eq:sep-nae-exponent} is subpolynomial in $\QE(F_k)$.

\subsection{A finite witness for the transfer principle}
\label{subsec:sep-balanced}

Consider $F_k\circ\OR_m$. Degree and adversary composition give
\[
 \deg(F_k\circ\OR_m)=2^k m,
 \qquad \Adv(F_k\circ\OR_m)=a^k\sqrt m.
\]
To minimize their maximum relative to the number $3^km$ of sensitive
variables, we balance the two expressions. Solving
$2^km=a^k\sqrt m$ gives
$m=(a/2)^{2k}=(9/8)^k$. Define
\begin{equation}\label{eq:sep-family}
 m_k=\left\lceil(9/8)^k\right\rceil,
 \qquad G_k=F_k\circ\OR_{m_k},\qquad N_k=3^k m_k.
\end{equation}
Rounding upward ensures that degree is at least the adversary bound.
Consequently,
\[
 \QElim(G_k)=2^km_k.
\]
Every raw input bit is sensitive at zero: flipping it changes its OR
output, and $F_k$ is sensitive to that changed outer bit. Hence
$s(G_k)=\bs(G_k)=\fbs(G_k)=N_k$. Since
$N_k=\Theta((27/8)^k)$ and $\QElim(G_k)=\Theta((9/4)^k)$,
we have $\fbs(G_k)=\Theta(\QElim(G_k)^{3/2})$.

For any fixed $\alpha<3/2$, choose an index $k_0$ large enough that
$g=G_{k_0}$ satisfies $\fbs(g)>\QElim(g)^\alpha$.
Corollary~\ref{cor:tr-four-way} rules out a universal exponent-$\alpha$
bound for $s$, $\bs$, or $\fbs$ in terms of $\QE$. Recursive powers of
this fixed $g$ supply a separating family.

The transfer theorem alone does not give a uniform quantitative bound
for the changing sequence $G_k$, since its fixed-gadget constants may
depend on $k$. For that purpose we apply the composition algorithm
directly.

\subsection{An explicit family attaining exponent \texorpdfstring{$3/2$}{3/2}}

\begin{theorem}[Exponent-$3/2$ separation]\label{thm:sep-three-halves}
Let $G_k$ and $N_k$ be as in~\eqref{eq:sep-family}. For every
$L\in\mathcal L=\{s,\bs,\fbs,\cert,D\}$,
\begin{equation}\label{eq:sep-measures}
 L(G_k)=N_k=\Theta((27/8)^k).
\end{equation}
Moreover,
\begin{align}
 \deg(G_k)=\QElim(G_k)&=2^k m_k=\Theta((9/4)^k),
       \label{eq:sep-rate}\\
 \tfrac12(9/4)^k\le\QE(G_k)
   &\le(9/4)^k\exp\!\left(O((\log(k+2))^2)\right).
       \label{eq:sep-exact}
\end{align}
Writing $q_k=\QE(G_k)$, we have
\begin{equation}\label{eq:sep-endpoint-loss}
 \frac{q_k^{3/2}}{\exp(O((\log\log(q_k+3))^2))}
 \le L(G_k)\le O(q_k^{3/2})
 \qquad(L\in\mathcal L).
\end{equation}
In particular, for each fixed $\alpha<3/2$ and all sufficiently large $k$,
\begin{equation}\label{eq:sep-divergence}
 L(G_k)\ge q_k^\alpha \qquad(L\in\mathcal L).
\end{equation}
More precisely, the same family satisfies
\[
 \lim_{k\to\infty}\frac{L(G_k)}{q_k^\alpha}=\infty
 \qquad(\alpha<3/2,\ L\in\mathcal L).
\]
\end{theorem}

\begin{proof}
\emph{Sensitivity and the exact-query lower bound.}
The full sensitivity at zero proved above gives
$s(G_k)=\bs(G_k)=\fbs(G_k)=N_k$. The inequalities
$s\le\cert\le D\le N_k$ give the two remaining equalities
in~\eqref{eq:sep-measures}. Also,
\[
 \deg(G_k)=2^km_k,\qquad \Adv(G_k)=a^k\sqrt{m_k}.
\]
Our choice $m_k\ge(a/2)^{2k}$ makes the first quantity larger, so
$\QElim(G_k)=2^km_k$. Since $m_k=\Theta((9/8)^k)$, this proves
\eqref{eq:sep-rate}. The polynomial method gives
\[
 \QE(G_k)\ge\tfrac12\deg(G_k)\ge\tfrac12(9/4)^k.
\]

\emph{An exact algorithm at the balanced rate.}
Apply Theorem~\ref{thm:degree-error} to the outer function $F_k$ and
the inner function $\OR_{m_k}$. The required inner error is
$1/(4\deg(F_k))=1/(4\cdot2^k)$. Quantum search and standard error
reduction~\cite{BHMT,BuhrmanDeWolf2002} give
\[
 Q_{1/(4\cdot2^k)}(\OR_{m_k})
 =O((k+1)\sqrt{m_k}).
\]
The composition bound and~\eqref{eq:sep-nae-query} now imply
\begin{align*}
 \QE(G_k)
 &\le O\!\left(2^km_k+
       (k+1)\sqrt{m_k}\,\QE(F_k)\right)\\
 &\le O\!\left(2^km_k+
       (k+1)a^k\sqrt{m_k}
       \exp\!\left(O((\log(k+2))^2)\right)\right).
\end{align*}
The balancing choice has made both principal costs equal at the
exponential scale:
\[
 2^km_k=\Theta((9/4)^k),\qquad
 a^k\sqrt{m_k}=\Theta((9/4)^k).
\]
Absorbing the factor $k+1$ into the stated overhead proves the upper
bound in~\eqref{eq:sep-exact}.

\emph{The separation bound.}
The identity $27/8=(9/4)^{3/2}$ and \eqref{eq:sep-exact} give
\[
 (9/4)^k=\Theta(L(G_k)^{2/3}),\qquad
 q_k\le O(L(G_k)^{2/3})\exp(O((\log(k+2))^2)).
\]
Also $\log(q_k+3)=\Theta(k+1)$ by \eqref{eq:sep-exact}. Raising the
upper bound to the power $3/2$ and rearranging gives the lower bound
in \eqref{eq:sep-endpoint-loss}. The lower bound on $q_k$ in
\eqref{eq:sep-exact} gives the other direction. For every $\alpha<3/2$,
the factor $q_k^{3/2-\alpha}$ eventually exceeds the denominator in
\eqref{eq:sep-endpoint-loss}, giving \eqref{eq:sep-divergence}.
Indeed, for each such $\alpha$,
\[
 \log\!\left(\frac{L(G_k)}{q_k^\alpha}\right)
 \ge (3/2-\alpha)\log q_k-O((\log\log(q_k+3))^2).
\]
Since $q_k$ tends to infinity, the right-hand side tends to infinity.
This proves the stated limit.
\end{proof}

The loss in \eqref{eq:sep-endpoint-loss} is
$(\log(q_k+3))^{O(\log\log(q_k+3))}$. Obtaining a fixed power of a
logarithm as the loss would require a sharper bound.

The family obtained by fixing a witness and then composing it can also
be described explicitly. For $g=G_{k_0}$ as above, we have $g(0)=0$
and full sensitivity at zero. Therefore
\[
 s(g^{\circ t})=\bs(g^{\circ t})=\fbs(g^{\circ t})=N_{k_0}^t,
 \qquad \QE(g^{\circ t})=\QElim(g)^t\exp(o(t)).
\]
These powers certify every chosen exponent below $3/2$.
For the fixed $\alpha$ used to choose $g$, we have
\[
 \lim_{t\to\infty}
 \frac{L(g^{\circ t})}{\QE(g^{\circ t})^\alpha}=\infty
 \qquad(L\in\{s,\bs,\fbs\}),
\]
because $N_{k_0}/\QElim(g)^\alpha>1$ and the exact cost has only a
subexponential overhead in $t$.
Theorem~\ref{thm:sep-three-halves} additionally controls the loss for the
original family $G_k$, whose balancing parameter changes with $k$.

\begin{corollary}[A composition counterexample from the same family]
\label{cor:sep-composition-gap}
The functions in~\eqref{eq:sep-family} satisfy
\begin{equation}\label{eq:sep-composition-gap}
 \QE(F_k\circ\OR_{m_k})
 \le Q(F_k)\QE(\OR_{m_k})
      \left(\frac{2\sqrt2}{3}\right)^k
      \exp\!\left(O((\log(k+2))^2)\right).
\end{equation}
Thus the multiplicative simulation cost exceeds
$\QE(F_k\circ\OR_{m_k})$ by a factor $\exp(\Omega(k))$.
In particular,
\[
 \lim_{k\to\infty}
 \frac{\QE(F_k\circ\OR_{m_k})}{\QE(F_k)\QE(\OR_{m_k})}=0.
\]
\end{corollary}

\begin{proof}
The standard bounds give $Q(F_k)=\Theta(a^k)$ and
$\QE(\OR_{m_k})=\Theta(m_k)$~\cite{BBCMW,Reichardt}.
Divide~\eqref{eq:sep-exact} by their product and use
$m_k\ge(9/8)^k$. The ratio of the exponential factors is
\[
 \frac{(9/4)^k}{a^k(9/8)^k}
 =\left(\frac2a\right)^k
 =\left(\frac{2\sqrt2}{3}\right)^k.
\]
Its base is smaller than one, so the exponential decay dominates the
$\exp(O((\log(k+2))^2))$ overhead. Finally,
$Q(F_k)\le\QE(F_k)$ gives the stated gap for exact simulation.
\end{proof}

\section{A nearly quadratic gap between randomized and exact quantum queries}
\label{sec:randomized-separation}

We next use the recursive algorithm underlying the transfer principles
to separate randomized query complexity from exact quantum query
complexity. For every fixed $0<\eps<1$, we construct total functions
with unbounded exact query complexity and $R\ge Q_E^{2-\eps}$.
We also give a single family with an explicit subpolynomial loss at
exponent two. This improves the previous
exponent-$3/2$ separation of Ambainis et al.~\cite{PointerFunctions},
which also follows from the cheat-sheet framework
\cite[Theorem~24]{ABKCheatSheets}.

The construction combines two known ingredients with our algorithm.
A cheat-sheet gadget has small degree and small bounded-error quantum
complexity. Sabotage complexity provides a large randomized lower bound
that composes without losing a constant at each level. Our recursive-rate
theorem then turns the two small parameters into an exact quantum upper
bound. For a fixed target exponent we choose the gadget first. For the
quantitative bound at exponent two we choose both its size and its depth.

\subsection{The recursive separation criterion}

We use randomized sabotage complexity, introduced by Ben-David and
Kothari~\cite{BKSabotage}. In its equivalent search formulation, an
algorithm receives a partial assignment compatible with both output
values of $u$ and must find an unspecified coordinate. The algorithm
has zero error; $\operatorname{RS}(u)$ is its minimum worst-case expected
query cost. The relevant lower bound and composition theorem are
\begin{equation}
\label{eq:sabotage-basic}
 R(u)\ge\frac{\operatorname{RS}(u)}3,
 \qquad
 \operatorname{RS}(u\circ v)
 \ge\operatorname{RS}(u)\operatorname{RS}(v)
 \qquad\text{\cite[Theorems~3.4 and~4.5]{BKSabotage}}.
\end{equation}
In particular,
$R(h^{\circ k})\ge\operatorname{RS}(h)^k/3$
\cite[Corollary~4.6]{BKSabotage}. We invoke these known results without
reproving them. Their coefficient-one composition bound is important:
a loss of a constant at every level would change the exponential base.

\begin{proposition}[A randomized separation from a fixed gadget]
\label{prop:rs-fixed-gadget}
Let $h$ be a fixed nonconstant total Boolean function, and set
\[
 r=\operatorname{RS}(h),
 \qquad
 \rho=\QElim(h)=\max\{\deg(h),\Adv(h)\}.
\]
Then
\begin{equation}
\label{eq:rs-fixed-gadget}
 R(h^{\circ k})\ge\frac{r^k}{3},
 \qquad
 \QE(h^{\circ k})=\rho^k\exp(o_h(k)).
\end{equation}
For every $\alpha>0$ with $r>\rho^\alpha$, there is $\eta>0$
such that, for all sufficiently large $k$,
\begin{equation}
\label{eq:rs-query-ratio}
 R(h^{\circ k})\ge e^{\eta k}\QE(h^{\circ k})^\alpha.
\end{equation}
In particular, $R(h^{\circ k})/\QE(h^{\circ k})^\alpha\to\infty$
as $k\to\infty$.
\end{proposition}

\begin{proof}
The first inequality in~\eqref{eq:rs-fixed-gadget} follows from sabotage
composition, and the second is Theorem~\ref{thm:recursive-growth-rate}.
Set $\eta=\tfrac12\log(r/\rho^\alpha)>0$. The factor
$r^k/\rho^{\alpha k}=e^{2\eta k}$ absorbs both the
$\exp(o_h(k))$ overhead and the constant $1/3$ for sufficiently large $k$.
This gives~\eqref{eq:rs-query-ratio}.
\end{proof}

This is the separating-family direction of the transfer argument from
Section~\ref{sec:tr-transfer}. The composable measure is now sabotage
complexity, which lower-bounds $R$, while the exact recursive rate
supplies the algorithmic upper bound.

\subsection{A standard cheat-sheet gadget}

Let $u_n=\AND_n\circ\OR_n$, viewed as a function on an $n\times n$
Boolean matrix. A certificate specifies either a zero row or one
nonzero entry in every row. Let $h_n$ be its standard cheat-sheet
function~\cite{ABKCheatSheets}: its input contains $O(\log n)$ copies
of the matrix and an array of certificate records. The copy outputs
specify an address, and $h_n$ accepts exactly when the record at that
address contains valid certificates agreeing with those outputs.
This defines a total Boolean function, including on inputs with invalid
records. We use the following established estimates.

\begin{lemma}[Known cheat-sheet estimates]
\label{lem:randomized-gadgets}
There are total Boolean functions $h_n$ on $v_n=n^{O(1)}$ bits with
\begin{equation}
\label{eq:randomized-gadget-bounds}
 n\le\deg(h_n)=\widetilde O(n),
 \qquad Q(h_n)=\widetilde O(n),
 \qquad \operatorname{RS}(h_n)=\widetilde\Omega(n^2).
\end{equation}
Consequently,
\begin{equation}
\label{eq:randomized-gadget-rate}
 n\le\rho_n:=\QElim(h_n)=\widetilde O(n),
 \qquad r_n:=\operatorname{RS}(h_n)=\widetilde\Omega(n^2).
\end{equation}
\end{lemma}

The upper bounds are the specialization of
\cite[Theorem~16 and Lemmas~17 and~19]{ABKCheatSheets} to
$Q(u_n)=O(n)$ and a deterministic certificate verifier of cost
$O(n\log n)$. For sabotage complexity, the cheat-sheet observation in
\cite[Section~7.1]{BKSabotage} gives
$\operatorname{RS}(u_{\mathrm{CS}})=\widetilde\Omega(R(u))$;
combine it with $R(u_n)=\Theta(n^2)$~\cite{BuhrmanDeWolf2002}.
These are inputs from the literature, rather than new cheat-sheet
bounds.

For clarity, the lower degree bound follows from a simple restriction.
Leave the first column of one matrix free and fix its remaining entries
to zero. In every other matrix, fix the first column to one. Fix the
record at the all-ones address to certify the first-column entries,
and make every other record invalid. The restricted function accepts
exactly when all $n$ free entries are one, so it is $\AND_n$.
Thus $\deg(h_n)\ge n$. Finally, the adversary characterization
$\Adv(h_n)=O(Q(h_n))$ gives~\eqref{eq:randomized-gadget-rate}.

Both upper bounds in~\eqref{eq:randomized-gadget-bounds} are needed:
our exact recursive algorithm is controlled by the larger of degree
and the adversary bound. Making only one of these parameters small
would not suffice.

\subsection{Approaching separation exponent two}

\begin{theorem}[A nearly quadratic randomized--exact quantum gap]
\label{thm:randomized-quadratic}
For every $0<\eps<1$ there is a fixed total Boolean function
$h=h_\eps$ with $\deg(h)\ge2$ such that, for all sufficiently large $k$,
\begin{equation}\label{eq:randomized-nearly-quadratic}
 R(h^{\circ k})\ge\QE(h^{\circ k})^{2-\eps}.
\end{equation}
For the same fixed gadget,
\[
 \lim_{k\to\infty}
 \frac{R(h^{\circ k})}{\QE(h^{\circ k})^{2-\eps}}=\infty.
\]
There is also a family of total Boolean functions $W_n$ with unbounded
exact query complexity satisfying
\begin{equation}\label{eq:query-quadratic-explicit}
 R(W_n)\ge
 \frac{\QE(W_n)^2}
 {\exp\!\left(O\!\left(\sqrt{\log(\QE(W_n)+2)
                        \log\log(\QE(W_n)+3)}\right)\right)}.
\end{equation}
In particular, this one family witnesses every fixed exponent
$\alpha<2$ in the sense that
\[
 \lim_{n\to\infty}\frac{R(W_n)}{\QE(W_n)^\alpha}=\infty.
\]
\end{theorem}

\begin{proof}
By Lemma~\ref{lem:randomized-gadgets}, there are constants $C_1,C_2>0$
and $a,b\ge0$ such that for all sufficiently large $n$,
\[
 r_n\ge\frac{C_1 n^2}{(\log n)^a},\qquad
 n\le \rho_n\le C_2 n(\log n)^b.
\]

\proofstep{A fixed gadget for each exponent.}
For fixed $\eps>0$, the bound
\[
 \frac{r_n}{\rho_n^{2-\eps}}\ge
 \frac{C_1}{C_2^{2-\eps}}\,
 \frac{n^\eps}{(\log n)^{a+b(2-\eps)}}
\]
exceeds one for sufficiently large $n$. Fix one such $n\ge2$ and apply
Proposition~\ref{prop:rs-fixed-gadget} to $h=h_n$ with $\alpha=2-\eps$.
Degree composition gives $\QE(h^{\circ k})\ge n^k/2$.
The exponential factor in \eqref{eq:rs-query-ratio} makes
$R(h^{\circ k})/\QE(h^{\circ k})^{2-\eps}$ tend to infinity as the
depth increases.

\proofstep{A quantitative bound with varying gadgets.}
Write $\delta_n=\deg(h_n)$ and let $v_n$ be the arity of $h_n$.
Choose an absolute $p$ such that $v_n\le n^p$ for sufficiently large
$n$. We need an exact algorithm estimate with constants uniform in
$n$. For $j\ge2$, apply Theorem~\ref{thm:degree-error} with outer
function $h_n^{\circ(j-1)}$ and inner function $h_n$. The correction
cost is $O(v_n\delta_n^{j-1})$. The outer degree is
$\delta_n^{j-1}$, so the required error reduction for $h_n$ costs
$O(j\log n)$ repetitions of a bounded-error algorithm. Here we use
$\delta_n\le v_n$ and $Q(h_n)=O(\Adv(h_n))=O(\rho_n)$.
For $u_j=\QE(h_n^{\circ j})/\rho_n^j$ this gives, with an absolute $C$,
\[
 u_j\le Cn^p+Cj\log n\,u_{j-1},\qquad u_1\le n^p.
\]
The bound on $u_1$ follows by reading the input. Induction yields
$u_j\le n^p(C'(j+1)\log n)^j$ for a sufficiently large absolute $C'$:
in the recurrence, the additive term is at most half the proposed
bound and the factor $Cj\log n$ times the preceding bound is at most
the other half.

Take $d_n=\lceil\log n/\log\log n\rceil$ for sufficiently large $n$
and put $W_n=h_n^{\circ d_n}$. Then
$d_n\log((d_n+1)\log n)=O(\log n)$ and
$d_n\log\log n=O(\log n)$. Since
$r_n\ge \rho_n^2\exp(-O(\log\log n))$, the uniform upper bound
and sabotage composition give
\begin{equation}\label{eq:quadratic-uniform-scales}
 \tfrac12 n^{d_n}\le\QE(W_n)\le\rho_n^{d_n}\exp(O(\log n)),
 \qquad
 R(W_n)\ge\rho_n^{2d_n}\exp(-O(\log n)).
\end{equation}
The lower bound on $\QE(W_n)$ uses degree composition.
In particular, $\log\QE(W_n)=\Theta(d_n\log n)$ and
$\log\log\QE(W_n)=\Theta(\log\log n)$. Thus
$\log n=O(\sqrt{\log\QE(W_n)\log\log\QE(W_n)})$.
Substituting this estimate in \eqref{eq:quadratic-uniform-scales}
gives \eqref{eq:query-quadratic-explicit}.
In particular, if $q_n=\QE(W_n)$, then for each fixed $\alpha<2$,
\[
 \log\!\left(\frac{R(W_n)}{q_n^\alpha}\right)
 \ge (2-\alpha)\log q_n
      -O\!\left(\sqrt{\log(q_n+2)\log\log(q_n+3)}\right).
\]
The loss term is $o(\log q_n)$, and $q_n$ tends to infinity by
\eqref{eq:quadratic-uniform-scales}. Thus the ratio tends to infinity
on the same family for every such $\alpha$.
\end{proof}

The loss in \eqref{eq:query-quadratic-explicit} is subpolynomial in
$\QE(W_n)$. Obtaining a quadratic separation with a fixed
polylogarithmic loss remains open for this construction.

\section{Different inner functions}
\label{sec:heterogeneous}

The subset expansion from Section~\ref{sec:algorithm} does not require
all inner functions to be identical. When block sizes differ, the
correction cost is determined by the blocks that can occur together in
an outer monomial. When the inner algorithms have different costs, their
error probabilities can be chosen separately. This section makes both
features explicit.

\subsection{Unequal blocks and weighted degree}
\label{subsec:sc-heterogeneous}
Let $f$ be a nonconstant total Boolean function on $n$ bits. For each
$i\in[n]$, let $g_i$ be a nonconstant total Boolean function on $m_i$
bits, and suppose that the input blocks are disjoint. We consider
\[
 F(x)=f\bigl(g_1(x^{(1)}),\ldots,g_n(x^{(n)})\bigr),
 \qquad N=\sum_{i=1}^n m_i.
\]
Write $H(y)=\sum_T\widehat H(T)y_T$ for the signed polynomial of $f$.
For positive weights $\mathbf w=(w_1,\ldots,w_n)$, define the
\emph{weighted degree}
\begin{equation}\label{eq:sc-weighted-degree-definition}
 D_{\mathbf w}(f)
 :=\max_{\widehat H(T)\ne0}\sum_{i\in T}w_i.
\end{equation}
Thus $D_{\mathbf m}(f)$ is the largest number of input bits contained in
the blocks indexed by a single monomial of $H$. This is the appropriate
correction cost: a nonzero derivative $\partial_SH$ can occur only when
$S$ is contained in the support of such a monomial. In particular,
$D_{\mathbf m}(f)\le\deg(f)\max_i m_i$, and the inequality can be strict.

We also recall the weighted form of the degree-composition identity
\cite[Lemma~3.1(1)]{Tal}. If $d_i=\deg(g_i)$, then
\begin{equation}\label{eq:sc-weighted-degree}
 \deg(F)=D_{\mathbf d}(f).
\end{equation}
This standard polynomial substitution identity will provide lower bounds
matching the algorithm in some cases.

\begin{proposition}[Exact evaluation with different inner functions]
\label{prop:sc-heterogeneous}
Choose $0<\eps_i<1/2$ with $\sum_i\eps_i\le1/4$, and put
$t_*:=\max_i Q_{\eps_i}(g_i)$. Then
\begin{equation}\label{eq:sc-heterogeneous}
 \QE(F)\le6D_{\mathbf m}(f)+12\QE(f)t_*.
\end{equation}
\end{proposition}
\begin{proof}
We use the same branches as in Theorem~\ref{thm:main}, but normalize the
residual for block $i$ by its own error bound. There are two quantities
to control: the maximum cost of a branch and the sum of the branch
weights.

For each $i$, choose an algorithm $A_i$ with error at most $\eps_i$,
and pad it to $t_*$ queries using the dummy query coordinate. Fix an
input $x$. Let $z_i=(-1)^{g_i(x^{(i)})}$ be the correct sign, let $b_i$
be the signed output bias of $A_i$, and write $e_i=z_i-b_i$. Then
$|e_i|\le2\eps_i$, and the exact subset expansion is
\begin{equation}\label{eq:sc-unequal-expansion}
 H(z)=\sum_{S:\partial_SH\ne0}e_S\,\partial_SH(b).
\end{equation}
As before, the quantities $b_i$ and $e_i$ are used to analyze the return
amplitudes; the algorithm does not first estimate them.

For a fixed set $S$, the derivative-correlation circuit of
Lemma~\ref{lem:derivative-circuit} uses $A_i$ whenever the outer query
register selects block $i$. The padding permits this selection
coherently, with one common query schedule. Its cost is at most
$4\QE(f)t_*$. Independently, Lemma~\ref{lem:residual-lookup} implements
the return amplitude $e_i/(2\eps_i)$ by reading and unreading block $i$,
at cost $2m_i$. Multiplying these return amplitudes therefore gives a
branch with return amplitude
\[
 a_S(x)=\partial_SH(b)\prod_{i\in S}\frac{e_i}{2\eps_i}.
\]
If $\partial_SH\ne0$, some set $T$ with $\widehat H(T)\ne0$ contains
$S$. Consequently,
\[
 \sum_{i\in S}m_i\le D_{\mathbf m}(f),
\]
and every branch uses at most
\[
 B=2D_{\mathbf m}(f)+4\QE(f)t_*
\]
queries. Padding these complete branches to the same schedule makes
their coherent selection cost $B$, irrespective of the number of
branches.

It remains to bound the normalization. Assign branch $S$ the weight
$w_S=\prod_{i\in S}2\eps_i$. Then
\[
 L:=\sum_{S:\partial_SH\ne0}w_S
 \le\prod_{i=1}^n(1+2\eps_i)
 \le\exp\!\left(2\sum_i\eps_i\right)
 \le e^{1/2}<2.
\]
Add a zero-return branch of weight $2-L$. The coherent selection
identity and~\eqref{eq:sc-unequal-expansion} now give return amplitude
\[
 \frac12\sum_S w_Sa_S(x)=\frac{H(z)}2.
\]
The same three-call exact amplification as in
Theorem~\ref{thm:main} produces the clean sign $H(z)$ at cost $3B$.
This is~\eqref{eq:sc-heterogeneous}.
\end{proof}

\subsection{Allocating the error budget}
Proposition~\ref{prop:sc-heterogeneous} separates the correction cost
from the choice of accuracies. Optimizing the latter gives
\begin{equation}\label{eq:sc-error-allocation}
 \QE(F)\le6D_{\mathbf m}(f)
 +12\QE(f)
 \min_{\substack{0<\eps_i<1/2\\\sum_i\eps_i\le1/4}}
       \max_i Q_{\eps_i}(g_i).
\end{equation}
The minimum is attained: the feasible maximum query costs form a
nonempty set of nonnegative integers. The choice $\eps_i=1/(4n)$
recovers a uniform accuracy requirement. A better allocation can give
the more expensive blocks larger error probabilities while keeping the
total error budget bounded.

For a convenient sufficient condition, put $q_i=Q(g_i)$. These numbers
are positive because each $g_i$ is nonconstant. Standard error
reduction gives $Q_\eps(g_i)=O(q_i\log(1/\eps))$. Hence a value
$\tau>0$ satisfying
\begin{equation}\label{eq:sc-common-budget}
 \sum_i\exp(-\tau/q_i)\le\frac14
\end{equation}
provides the feasible allocation $\eps_i=\exp(-\tau/q_i)$ and the bound
\[
 \QE(F)=O\!\left(D_{\mathbf m}(f)+\QE(f)\tau\right).
\]
The role of $\tau$ is to give every inner algorithm the same query
budget, rather than the same error probability.

For example, suppose a group of $r$ blocks has $q_i\le q_H$ and the
remaining $n-r$ blocks have $q_i\le q_L$, where $1\le r<n$. Taking
\[
 \tau=\max\{q_H\log(8r),\ q_L\log(8(n-r))\}
\]
makes the contribution of each group to
\eqref{eq:sc-common-budget} at most $1/8$. If the expensive blocks are
few and the remaining blocks are sufficiently inexpensive, this can
improve the $O(q_H\log n)$ budget obtained from uniform error reduction.

\subsection{Unequal search blocks}
Suppose now that each $g_i$ is either $\OR_{m_i}$ or $\AND_{m_i}$.
Since $d_i=m_i$, the degree identity~\eqref{eq:sc-weighted-degree} and
the polynomial method~\cite{BBCMW} give
\[
 \QE(F)\ge\frac12D_{\mathbf m}(f).
\]
The standard small-error search bound~\cite{BCWZ,deWolf} states that
\[
 Q_{\eps_i}(g_i)
 =\Theta\!\left(\min\{m_i,\sqrt{m_i\log(1/\eps_i)}\}\right)
 \qquad(0<\eps_i\le1/4).
\]
Every feasible allocation in
Proposition~\ref{prop:sc-heterogeneous} has $\eps_i\le1/4$, so it gives
\[
 t_*=O\!\left(\max_i
       \min\{m_i,\sqrt{m_i\log(1/\eps_i)}\}\right).
\]
Thus the approximate-evaluation term is the only possible source of a
gap between the upper and lower bounds. Whenever the allocation
satisfies $\QE(f)t_*=O(D_{\mathbf m}(f))$, they match:
\begin{equation}\label{eq:sc-unequal-tight}
 \QE(F)=\Theta(D_{\mathbf m}(f)).
\end{equation}
This criterion accommodates unequal block sizes without replacing all
of them by the largest size.
\section{Applications to quantum communication}
\label{sec:communication}

We now apply the query algorithms to two communication constructions.
For the explicit recursive NAE families, a coordinatewise AND gadget
preserves a disjointness restriction, which gives a linear randomized
lower bound. For the nearly quadratic separation, we instead use an
inner-product gadget and a randomized lifting theorem. Both
constructions use the standard exact query-to-communication simulation
for their quantum upper bounds.

Throughout this section, $Q_E^{\mathrm{cc}}$ denotes exact quantum
communication complexity without prior entanglement, and
$R_{1/3}^{\mathrm{cc}}$ denotes public-coin randomized communication
complexity with worst-case error at most $1/3$.
Protocols may have unrestricted interaction. Local computation is free,
and communication costs are worst-case costs.

\subsection{Standard simulation and lower bounds}
\label{subsec:comm-simulation}
For a Boolean function $h$ on $N$ bits, define its coordinatewise-AND
communication lift by
\[
 \mathcal H_h(x,y)=h(x_1\land y_1,\ldots,x_N\land y_N),
 \qquad x,y\in\{0,1\}^N.
\]
Alice receives $x$ and Bob receives $y$. We will use three established
facts: query simulation for the quantum upper bound, disjointness for
the randomized lower bound, and a rank identity for an exact quantum
lower bound.

\begin{lemma}[Query simulation; Buhrman, Cleve, and Wigderson~\cite{BCW98}]
\label{lem:comm-simulation}
For every Boolean function $h$ on $N$ bits,
\begin{equation}\label{eq:comm-simulation}
 Q_E^{\mathrm{cc}}(\mathcal H_h)
 =O\!\left(\log(N+1)Q_E(h)\right).
\end{equation}
The protocol requires no prior entanglement.
\end{lemma}

The simulation preserves the error of the query algorithm, so an exact
query algorithm gives an exact communication protocol. We invoke the
simulation without reproving it.

For the classical lower bound, recall Razborov's $\Omega(N)$ randomized
lower bound for disjointness even under the promise that the intersection
has size zero or one~\cite{Razborov92}. If $h(0^N)=0$ and $h(e_i)=1$
for every unit vector $e_i$, then restricting the communication inputs
to this promise makes $\mathcal H_h$ distinguish these two cases.
Together with the trivial $O(N)$ deterministic upper bound, this gives
\begin{equation}\label{eq:comm-classical}
 R_{1/3}^{\mathrm{cc}}(\mathcal H_h)=\Theta(N),
 \qquad D^{\mathrm{cc}}(\mathcal H_h)=\Theta(N).
\end{equation}
Thus full sensitivity at zero gives a direct classical lower bound for
the AND lift.

The heterogeneous algorithm also has an immediate communication
consequence. For $F=f(g_1,\ldots,g_n)$ on $N=\sum_i m_i$ input bits,
Proposition~\ref{prop:sc-heterogeneous} and
Lemma~\ref{lem:comm-simulation} imply
\begin{equation}\label{eq:comm-weighted}
 \begin{split}
 Q_E^{\mathrm{cc}}(\mathcal H_F)
 =O\!\Bigl(&\log(N+1)
       \bigl[D_{\mathbf m}(f)
          +Q_E(f)\max_iQ_{\eps_i}(g_i)\bigr]\Bigr),\\
 &\hspace{35mm}\sum_i\eps_i\le\frac14 .
 \end{split}
\end{equation}
The error allocation can therefore be optimized exactly as in
\eqref{eq:sc-error-allocation}. Moreover, if $f$ and each $g_i$ are zero
at zero and one at every unit vector, then $F$ has the same property.
In this case~\eqref{eq:comm-classical} gives randomized communication
complexity $\Theta(N)$.

To obtain an exact quantum lower bound, let $\operatorname{mon}(h)$
denote the number of nonzero monomials in the multilinear representation
of $h$ in $\{0,1\}$ variables.

\begin{lemma}[AND-lift rank and exact quantum log-rank~\cite{BWComm}]
\label{lem:comm-rank}
For the real Boolean communication matrix
$M_h=(h(x\land y))_{x,y}$,
\[
 \operatorname{rank}(M_h)=\operatorname{mon}(h).
\]
Consequently,
\begin{equation}\label{eq:comm-rank-lower}
 Q_E^{\mathrm{cc}}(\mathcal H_h)
 \ge\tfrac12\log_2\operatorname{mon}(h).
\end{equation}
\end{lemma}

The rank identity and the exact quantum log-rank bound are,
respectively, Corollary~6 and Theorem~2 of Buhrman and de
Wolf~\cite{BWComm}. We use these established results without proof.
Their unassisted consequence suffices for the communication model
considered here.

\subsection{Recursive NAE and the balanced family}
\label{subsec:comm-recursive}
Recall the families
\[
 F_k=\NAE_3^{\circ k},\qquad
 m_k=\lceil(9/8)^k\rceil,\qquad
 G_k=F_k\circ\OR_{m_k},\qquad N_k=3^km_k.
\]
Both are zero at zero and one at every unit vector. Their AND lifts
therefore have linear randomized complexity. The purpose of the next
theorem is to quantify the exact quantum savings and, for the balanced
family, prove a matching lower exponent.

\begin{theorem}[Exact protocols for the recursive families]
\label{thm:communication}
Let $a=3/\sqrt2$. For $F_k$ on $N=3^k$ bits,
\begin{equation}\label{eq:comm-nae}
 \begin{split}
 Q_E^{\mathrm{cc}}(\mathcal H_{F_k})
 &\le N^{\log_3 a}
       \exp\!\left(O((\log\log N)^2)\right),\\
 R_{1/3}^{\mathrm{cc}}(\mathcal H_{F_k})&=\Theta(N).
 \end{split}
\end{equation}
For $G_k$ on $N=N_k$ bits,
\begin{equation}\label{eq:comm-balanced}
 \Omega(N^{2/3})
 \le Q_E^{\mathrm{cc}}(\mathcal H_{G_k})
 \le N^{2/3}\exp\!\left(O((\log\log N)^2)\right),
\end{equation}
and $R_{1/3}^{\mathrm{cc}}(\mathcal H_{G_k})=\Theta(N)$.
In both cases $D^{\mathrm{cc}}=\Theta(N)$. Writing
$q_k=Q_E^{\mathrm{cc}}(\mathcal H_{G_k})$, the balanced family satisfies
\begin{equation}\label{eq:comm-balanced-separation}
 \frac{q_k^{3/2}}{\exp(O((\log\log(q_k+3))^2))}
 \le R_{1/3}^{\mathrm{cc}}(\mathcal H_{G_k})
 \le O(q_k^{3/2}).
\end{equation}
\end{theorem}
\begin{proof}
The classical claims follow from the sensitivity calculation in
Section~\ref{sec:sep-separations} and~\eqref{eq:comm-classical}.
We first establish the quantum upper bounds, and then count monomials
to obtain the lower bound for $G_k$.

By Theorem~\ref{thm:nae-near-tight} and query simulation,
\[
 Q_E^{\mathrm{cc}}(\mathcal H_{F_k})
 \le O(k)a^k\exp\!\left(O((\log(k+2))^2)\right).
\]
Since $N=3^k$, we have $a^k=N^{\log_3a}$ and $k=\Theta(\log N)$.
The factor $O(k)$ is absorbed by the displayed subpolynomial overhead,
proving~\eqref{eq:comm-nae}. Applying the same simulation to
Theorem~\ref{thm:sep-three-halves} gives
\[
 Q_E^{\mathrm{cc}}(\mathcal H_{G_k})
 \le(9/4)^k\exp\!\left(O((\log(k+2))^2)\right).
\]
Here the additional simulation factor $O(k)$ has again been absorbed.
The identities
\[
 N_k=\Theta((27/8)^k),\qquad (27/8)^{2/3}=9/4
\]
give the upper bound in~\eqref{eq:comm-balanced}.

For the lower bound, fix $m\ge2$. Let $s_j$ count the nonzero monomials
in the $\{0,1\}$ polynomial of $F_j\circ\OR_m$, with $F_0$ the identity
function. The initial polynomial
\[
 \OR_m(z)=1-\prod_{i=1}^m(1-z_i)
\]
has $s_0=2^m-1$ monomials and no constant term. The polynomial of the
ternary gate is
\[
 \NAE_3(u,v,w)=u+v+w-uv-uw-vw.
\]
At the next recursive level, the three child polynomials have disjoint
variable sets and zero constant term. A monomial from a linear term is
supported in exactly one child; a monomial from a quadratic term is
supported in exactly two children. These two types cannot cancel.
Within a fixed child or pair of children, the support uniquely
determines the constituent monomials, so cancellation cannot occur there
either. We therefore obtain the exact recurrence
\[
 s_{j+1}=3s_j+3s_j^2.
\]
For a lower bound, it is enough to retain the inequality
$s_{j+1}\ge s_j^2$. Iterating it gives
\[
 \log_2s_k\ge2^k\log_2(2^m-1)\ge2^{k-1}m.
\]
Lemma~\ref{lem:comm-rank} now yields
\[
 Q_E^{\mathrm{cc}}(\mathcal H_{F_k\circ\OR_m})
 \ge2^{k-2}m.
\]
Finally, set $m=m_k$. Since
$2^km_k=\Theta((9/4)^k)=\Theta(N_k^{2/3})$, this is the lower bound
in~\eqref{eq:comm-balanced}. Since $\log q_k=\Theta(\log N)$,
raising the bounds in~\eqref{eq:comm-balanced} to the power $3/2$
and using the linear randomized cost gives
\eqref{eq:comm-balanced-separation}.
\end{proof}

For the same recursive NAE communication function, Ambainis's
STOC~2013 algorithm~\cite[Section~3.6]{AmbExact} used
$O(N^{0.8675\ldots}\log N)$ qubits.
Equation~\eqref{eq:comm-nae} reduces the exponent to
$\log_3(3/\sqrt2)=0.684535\ldots$. Adding the OR layer yields the
explicit exponent-$3/2$ separation in~\eqref{eq:comm-balanced};
the monomial count certifies the matching exponent for its exact
quantum communication complexity.

\subsection{A nearly quadratic randomized--exact separation}
\label{subsec:comm-quadratic}
The functions from Theorem~\ref{thm:randomized-quadratic} need not be
fully sensitive at zero. We therefore cannot use the preceding
disjointness restriction to transfer their randomized query lower
bounds. Instead, we use the inner-product gadget
\[
 \operatorname{IP}_b(u,v)=\bigoplus_{j=1}^b u_jv_j,
 \qquad u,v\in\{0,1\}^b.
\]
For an $N$-bit Boolean function $f$, define
\begin{equation}\label{eq:comm-ip-definition}
 \mathcal I_{f,b}(x,y)
 =f\bigl(\operatorname{IP}_b(x_1,y_1),\ldots,
         \operatorname{IP}_b(x_N,y_N)\bigr),
 \qquad x,y\in(\{0,1\}^b)^N.
\end{equation}
Choosing $b$ logarithmic in $N$ has two advantages: randomized lifting
preserves the query lower bound up to this logarithmic factor, and exact
query simulation incurs only a logarithmic overhead.

\begin{lemma}[Randomized lifting and exact simulation
 {\cite[Theorem~1.2]{CFKMP2021},\cite[Theorem~2.1]{BCW98}}]
\label{lem:comm-ip-lifting}
There is an absolute constant $c>0$ such that, for
$b=\lceil c\log_2(N+1)\rceil$ and every nonconstant total Boolean
function $f$ on $N$ bits,
\begin{align}
 R_{1/3}^{\mathrm{cc}}(\mathcal I_{f,b})
   &=\Theta\!\left(bR(f)\right),
       \label{eq:comm-ip-randomized}\\
 Q_E^{\mathrm{cc}}(\mathcal I_{f,b})
   &=O\!\left((\log(N+1)+b)\QE(f)\right).
       \label{eq:comm-ip-exact}
\end{align}
The exact quantum protocol requires no prior entanglement.
\end{lemma}

We record how the cited results specialize to this statement. Inner
product has uniform discrepancy at most $2^{-b/2}$, and hence satisfies
the low-discrepancy hypothesis of the lifting theorem. Its small error
increase is absorbed by constant-factor error reduction. The theorem
also permits a constant additive loss in the outer query lower bound;
when $R(f)$ is bounded, a sensitive-edge restriction to inner product
supplies the necessary $\Omega(b)$ bound. These standard
specializations give the constant-error formulation above. No new
lifting or simulation theorem is required.

We will also use the following lower bound for $b\ge2$:
\begin{equation}\label{eq:comm-ip-rank-degree}
 Q_E^{\mathrm{cc}}(\mathcal I_{f,b})
 \ge \tfrac12\deg(f)\log_2(2^b-1).
\end{equation}
To see this, regard $\mathcal I_{f,b}$ as the coordinatewise AND lift
of $f\circ\PAR_b$. Take a monomial of $f$ with support $S$ of size
$\deg(f)$. The polynomial of $\PAR_b$ on $\{0,1\}^b$ has a
nonzero term for every nonempty subset and zero constant term.
The substituted monomial therefore yields $(2^b-1)^{|S|}$
distinct monomials whose sets of nonempty blocks are exactly $S$.
Only the monomial of $f$ with support $S$ can contribute these terms.
The rank identity and exact log-rank bound in
Lemma~\ref{lem:comm-rank} give \eqref{eq:comm-ip-rank-degree}.

\begin{theorem}[Nearly quadratic communication separation]
\label{thm:communication-quadratic}
For every $0<\eps<1$, there is a family of total Boolean communication
functions $\mathcal F_k$ with unbounded exact quantum complexity and
\begin{equation}\label{eq:comm-quadratic-separation}
 R_{1/3}^{\mathrm{cc}}(\mathcal F_k)
 \ge \bigl(Q_E^{\mathrm{cc}}(\mathcal F_k)\bigr)^{2-\eps}
\end{equation}
for all sufficiently large $k$.
There is also a family $\mathcal W_n$ of total Boolean communication
functions with unbounded exact quantum complexity satisfying
\begin{equation}\label{eq:comm-quadratic-explicit}
 R_{1/3}^{\mathrm{cc}}(\mathcal W_n)
 \ge \frac{q_n^2}{\exp(O(\sqrt{\log(q_n+2)\log\log(q_n+3)}))},
 \qquad q_n=Q_E^{\mathrm{cc}}(\mathcal W_n).
\end{equation}
The classical protocols may use public randomness and unrestricted
interaction. The exact quantum upper bounds use no prior entanglement.
\end{theorem}

\begin{proof}
\proofstep{A fixed exponent.}
Apply Theorem~\ref{thm:randomized-quadratic} with $\eps/2$ to obtain
$f_k=h^{\circ k}$ on $N_k$ bits. Put
$b_k=\lceil c\log_2(N_k+1)\rceil$ as in Lemma~\ref{lem:comm-ip-lifting},
and set $\mathcal F_k=\mathcal I_{f_k,b_k}$.
The lemma gives
\begin{equation}\label{eq:comm-quadratic-costs}
 R_{1/3}^{\mathrm{cc}}(\mathcal F_k)
 \ge\Omega(b_k\QE(f_k)^{2-\eps/2}),\qquad
 Q_E^{\mathrm{cc}}(\mathcal F_k)\le O(b_k\QE(f_k)).
\end{equation}
The factor $\QE(f_k)^{\eps/2}$ grows exponentially in $k$ because
$\deg(h)\ge2$. It therefore absorbs the polynomial loss from
$b_k=\Theta_h(k)$ and the constants in
\eqref{eq:comm-quadratic-costs}, yielding
\eqref{eq:comm-quadratic-separation}.
By \eqref{eq:comm-ip-rank-degree}, the exact communication cost is
at least $\Omega(b_k\deg(h)^k)$.

\proofstep{A quantitative bound at exponent two.}
Use $W_n=h_n^{\circ d_n}$ from Theorem~\ref{thm:randomized-quadratic},
and let $\mathcal W_n=\mathcal I_{W_n,b_n}$ with
$b_n=\lceil c\log_2(v_n^{d_n}+1)\rceil$.
As $n\le v_n\le n^p$, the block length is
$b_n=\Theta(d_n\log n)$. Applying
Lemma~\ref{lem:comm-ip-lifting} to \eqref{eq:quadratic-uniform-scales}
gives
\[
 R_{1/3}^{\mathrm{cc}}(\mathcal W_n)
 \ge b_n\rho_n^{2d_n}\exp(-O(\log n)),\qquad
 q_n\le b_n\rho_n^{d_n}\exp(O(\log n)).
\]
The factor $b_n$ is polynomial in $\log n$, so these estimates give
$R_{1/3}^{\mathrm{cc}}(\mathcal W_n)\ge q_n^2\exp(-O(\log n))$.
Equation~\eqref{eq:comm-ip-rank-degree} gives
$q_n=\Omega(b_n n^{d_n})$. Together with the upper bound, this
yields $\log q_n=\Theta(d_n\log n)$ and hence
$\log n=O(\sqrt{\log q_n\log\log q_n})$.
This proves \eqref{eq:comm-quadratic-explicit}.
\end{proof}

\subsection{Comparison with earlier work}
\label{subsec:comm-literature}
Anshu et al.~\cite[Theorem~2]{ABBDGJKLS} established a separation of
exponent $3/2$ between randomized and exact quantum communication
complexity for total functions, improving Ambainis's earlier exponent
of approximately $1.15$~\cite{AmbExact}. Their exact quantum upper
bound also requires no prior entanglement.
Theorem~\ref{thm:communication-quadratic} attains every fixed exponent
below two. The new ingredient is the exact query algorithm for the
recursive cheat-sheet family; randomized lifting transfers its
classical lower bound, and exact query simulation transfers its
quantum upper bound.

For the balanced family, \eqref{eq:comm-balanced-separation} gives
the exponent-$3/2$ comparison with an explicit loss and a matching
upper exponent. For the cheat-sheet family,
\eqref{eq:comm-quadratic-explicit} quantifies the loss at exponent two.

\section{Conclusion and open problems}
\label{sec:conclusion}

The composition algorithm uses polynomial degree to bound the cost of
correcting approximate inner computations. Its degree-dependent error
threshold leads to exact algorithms for recursive functions whose
exponential rate is determined by degree and the general adversary bound.
The transfer principles then turn inequalities for individual gadgets
into universal exact-query bounds, while the explicit constructions
separate exact quantum complexity from sensitivity measures and from
randomized complexity.

Two sources of loss remain in the algorithms. First, the exact correction
reads selected inner blocks in full. Second, balancing the recursive
costs introduces an overhead beyond the optimal exponential rate.
Understanding whether the first loss can be removed would also clarify
the second.

\paragraph{Implementing the correction through an exact inner algorithm.}
Both composition proofs pay $m$ queries for each selected inner block.
It is natural to ask whether this cost can be replaced by $Q_E(g)$.

\begin{conjecture}[Composition with an exact inner correction]
\label{conj:arity-free}
There is an absolute constant $C$ such that every pair of nonconstant
total Boolean functions $f$ and $g$ satisfies
\[
 Q_E(f\circ g)\le C\left(\deg(f)Q_E(g)
       +Q_E(f)Q_{1/(4\deg(f))}(g)\right).
\]
\end{conjecture}

The obstacle is specific. In the first construction, the normalized
residual depends on the error probability of the approximate inner
algorithm. In the interpolation construction, the correction operators
also depend on its error state. An exact algorithm for $g$ supplies the
correct Boolean answer, but this does not immediately implement either
of these corrections with $O(Q_E(g))$ queries. A proof of the conjecture
must implement the required correction more efficiently or replace it
with a different construction.

\paragraph{Removing the recursive overhead.}
For a fixed gadget $h$, we proved
$Q_E(h^{\circ k})=\widetilde O_h(R(h^{\circ k})^2)$.
The logarithmic factor is unnecessary when
$\deg(h)<\Adv(h)^2$ or when $\deg(h)$ equals the arity of $h$.
Removing it in the remaining parameter regime is open. A uniform bound
$Q_E(f)=O(R(f)^2)$ for individual total functions would require an
argument beyond the fixed-gadget rate identity.

The simplest concrete question concerns recursive NAE.
\begin{problem}[Tight exact complexity of recursive NAE]
\label{prob:tight-nae}
Does $Q_E(\NAE_3^{\circ k})=\Theta((3/\sqrt2)^k)$ hold?
\end{problem}

Conjecture~\ref{conj:arity-free} would give a positive answer. We spell out
this implication because it isolates a recurrence that could also be
proved directly. Put $F_j=\NAE_3^{\circ j}$,
$E_j=Q_E(F_j)$, and $a=3/\sqrt2$. Fix an outer depth $b$ and apply the
conjecture to $F_b\circ F_\ell$. Since $\deg(F_b)=2^b$ and the target
inner error is constant once $b$ is fixed, adversary composition and
error reduction give
\[
 E_{\ell+b}\le C2^bE_\ell+B_ba^\ell,
\]
where $B_b$ is independent of $\ell$. Choose $b$ large enough that
$C2^b<a^b$, which is possible because $2<a$. For the normalized cost
$u_j=E_j/a^j$, the recurrence becomes
\[
 u_{\ell+b}\le\theta u_\ell+K_b,
 \qquad \theta=\frac{C2^b}{a^b}<1,
 \qquad K_b=\frac{B_b}{a^b}.
\]
Iterating on each residue class modulo $b$ bounds $u_j$ by a geometric
series. This gives $E_j=O(a^j)$, matching the adversary lower bound.
Thus proving the displayed fixed-depth recurrence would already settle
the question for NAE, without resolving the full composition conjecture.

\paragraph{The sensitivity exponent and its endpoint.}
\begin{problem}[Improving the sensitivity separation]
Can one find a family with $s(f)=Q_E(f)^{\alpha-o(1)}$ for some
$\alpha>3/2$? Can the present family attain
$s(f)=\widetilde\Omega(Q_E(f)^{3/2})$, or even
$s(f)=\Omega(Q_E(f)^{3/2})$?
\end{problem}

The first question asks for a larger exponent; the second asks whether
the overhead in the current exponent-$3/2$ construction can be removed.
The transfer theorem gives a finite-function formulation of the first
question's universal-bound counterpart:
\[
 s(F)=O(Q_E(F)^{3/2})\text{ for all }F
 \quad\Longleftrightarrow\quad
 \fbs(f)\le\max\{\deg(f),\Adv(f)\}^{3/2}\text{ for all }f.
\]
A gadget violating the right-hand inequality would therefore yield a
stronger asymptotic separation. A sharper quantitative upper bound on
the exact algorithm for $G_k$ could improve the loss in
\eqref{eq:sep-endpoint-loss}. For the randomized separations, an analogous
question is whether a fixed power of a logarithm can replace the
denominators in \eqref{eq:query-quadratic-explicit} and
\eqref{eq:comm-quadratic-explicit}.

{\small
\bibliographystyle{abbrvnat}
\bibliography{composition_rewritten_20261105}
}
\end{document}